\documentclass{iosart2c}

\usepackage{amsmath}
\usepackage{changepage}

\usepackage{subcaption}
\usepackage{fancyvrb,fancybox,calc} 
\usepackage[utf8]{inputenc}
\usepackage{csquotes}
\usepackage{microtype}
\newcommand{\cmark}{\ding{51}}%
\usepackage{comment}
\usepackage{wasysym}
\usepackage{numprint}
\usepackage{balance}
\usepackage{todonotes}

\usepackage{amsthm}
\usepackage[T1]{fontenc}
\usepackage{times}
\usepackage{listings}
\usepackage{tabularx}
\usepackage{graphicx}
\newtheorem{definition}{Definition}
\usepackage{adjustbox}
\usepackage{amssymb}
\usepackage{pdflscape}
\usepackage{paralist}
\usepackage{url}
\usepackage{hyperref}
\usepackage{cleveref}

\usepackage{natbib}
\usepackage{floatrow}

\usepackage{graphics}
\usepackage{epstopdf}
\usepackage{tikz}
\usepackage{pgf-pie}
\usepackage{pgfplots}
\pgfplotsset{compat=1.5}
\usepackage{textcomp}
\usepackage{enumitem}
\usepgfplotslibrary{dateplot}
\usepackage{todonotes}
\usepackage{pifont}
\usepackage{multirow}
\usepackage{colortbl}
\definecolor{gray}{gray}{0.8}
\definecolor{LightCyan}{rgb}{0.8,1,1}
\usepackage{comment}
\newcolumntype{d}[1]{D{.}{.}{#1}}
\usepackage{xcolor}
\definecolor{gray}{gray}{0.8}
\usepackage{booktabs}
\usepackage{wasysym}
\usepackage{amsmath}
\newcommand{\dss}{\ensuremath{\mathcal{D}}}

\newcommand{\bound}{\ensuremath{b}}

\usepackage[scaled=0.98]{couriers}

\lstdefinestyle{base}{
  language=sparql,
  emptylines=1,
  breaklines=true,
  basicstyle=\ttfamily\color{black},
  moredelim=**[is][\color{red}]{@}{@}
}

\definecolor{Gray}{gray}{0.85}
\definecolor{LightCyan}{rgb}{0.88,1,1}
\definecolor{LightCyan2}{rgb}{0.5,1,1}

\newcolumntype{a}{>{\columncolor{Gray}}c}
\newcolumntype{b}{>{\columncolor{white}}c}
\newcommand\mydots{\hbox to 1em{.\hss.\hss.}}

\definecolor{MyLightGray}{RGB}{200, 200,200}
\lstdefinelanguage{turtle}
{
    columns=fullflexible,
    keywordstyle=\color{red},
    morekeywords={PREFIX,SELECT,DISTINCT,UNION,FILTER,ORDER,BY,REGEX,STR,isBlank},
    morecomment=[l]{\#},
    tabsize=4,
    frame=lines,
    alsoletter={-?}, 
    morecomment=[s][\color{blue}]{<}{>},
    basicstyle=\scriptsize\ttfamily\color{black},
    morestring=[b][\color{black}]\",
    backgroundcolor=\color{background},    
}

\colorlet{punct}{red!60!black}
\definecolor{background}{HTML}{EEEEEE}
\definecolor{delim}{RGB}{20,105,176}
\colorlet{numb}{magenta!60!black}

\lstdefinelanguage{json}{
    basicstyle=\scriptsize\ttfamily,
    stepnumber=1,
    numbersep=8pt,
    showstringspaces=false,
    breaklines=true,
    frame=lines,
    backgroundcolor=\color{background},
    literate=
     *{0}{{{\color{numb}0}}}{1}
      {1}{{{\color{numb}1}}}{1}
      {2}{{{\color{numb}2}}}{1}
      {3}{{{\color{numb}3}}}{1}
      {4}{{{\color{numb}4}}}{1}
      {5}{{{\color{numb}5}}}{1}
      {6}{{{\color{numb}6}}}{1}
      {7}{{{\color{numb}7}}}{1}
      {8}{{{\color{numb}8}}}{1}
      {9}{{{\color{numb}9}}}{1}
      {:}{{{\color{punct}{:}}}}{1}
      {,}{{{\color{punct}{,}}}}{1}
      {\{}{{{\color{delim}{\{}}}}{1}
      {\}}{{{\color{delim}{\}}}}}{1}
      {[}{{{\color{delim}{[}}}}{1}
      {]}{{{\color{delim}{]}}}}{1},
}

\firstpage{1} \lastpage{5} \volume{1} \pubyear{2020}

\begin{document}

\begin{frontmatter}                        

%
\title{An Empirical Evaluation of Cost-based Federated SPARQL Query Processing Engines}


\runningtitle{An Empirical Evaluation of Cost-based Federated SPARQL Query Processing Engines}

\review{Ruben Verborgh, Ghent University, Belgium}{Stasinos Konstantopoulos, Institute and Informatics and Telecommunications, Greece; Five anonymous reviewers}{}

\author[C]{\fnms{Umair} \snm{Qudus}},
\author[A]{\fnms{Muhammad} \snm{Saleem}} 
,
\author[B]{\fnms{Axel-Cyrille} \snm{Ngonga Ngomo}}
and
\author[C]{\fnms{Young-koo} \snm{Lee}\thanks{Corresponding author. E-mail: yklee@khu.ac.kr}}

\address[C]{DKE, Kyung Hee University, South Korea\\
E-mail:\{umair.qudus,yklee\}@khu.ac.kr}
\address[A]{AKSW, Leipzig, Germany\\
E-mail: \{lastname\}@informatik.uni-leipzig.de}
\address[B]{University of Paderborn, Germany\\
E-mail: axel.ngonga@upb.de}

\begin{abstract}
Finding a good query plan is key to the optimization of query runtime. This holds in particular for cost-based federation engines, which make use of cardinality estimations to achieve this goal. A number of studies compare SPARQL federation engines across different performance metrics, including query runtime, result set completeness and correctness, number of sources selected and number of requests sent. Albeit informative, these metrics are generic and unable to quantify and evaluate the accuracy of the cardinality estimators of cost-based federation engines. To thoroughly evaluate cost-based federation engines, the effect of  estimated cardinality errors on the overall query runtime performance must be measured. In this paper, we address this challenge by presenting novel evaluation metrics targeted at a fine-grained benchmarking of cost-based federated SPARQL query engines. We evaluate five cost-based federated SPARQL query engines using existing as well as novel evaluation metrics by using LargeRDFBench queries. Our results provide a detailed analysis of the experimental outcomes that reveal novel insights, useful for the development of future cost-based federated SPARQL query processing engines.   

\end{abstract}


\end{frontmatter}

\section{Introduction}
The availability of increasing amounts of data published in RDF has led to the genesis of many federated SPARQL query engines. These engines vary widely in their approaches to generating a good query plan \cite{fineGrainEval2014,QOFSE2013,rdfquerysurvey2018,survey2000}. In general, there exist several possible plans that a federation engine can consider when executing a given query. These plans have different costs in terms of the resources required and the overall query execution time. Selection of the best possible plan with minimum cost is hence of key importance when devising cost-based federation engines; a fact which is corroborated by a plethora of works in database research \cite{howgood2015,qerror2009}.

In SPARQL query federation, index-free (heuristics-based) \cite{fedx2011,montoya2012heuristic,exSPARQLWeb2009} and index-assisted (cost-based) \cite{costfed2018,LHD2013,darq2008,Odyssey2017,SPLENDID2019,semagrow2015,hasnain2012cataloguing,hasnain2014roadmap,sihjoin2011,aderis2011,partitioned2012} engines are most commonly used for federated query processing \cite{fineGrainEval2014}. The heuristics-based federation engines do not store any pre-computed statistics and hence mostly use different heuristics to optimize their query plans \cite{fedx2011}. Cost-based engines make use of an index with pre-computed statistics about the datasets \cite{fineGrainEval2014}. Using cardinality estimates as principal input, such engines make use of cost models to calculate the cost of different query joins and generate optimized query plans. 
Most state-of-the-art cost-based federated SPARQL processing engines \cite{costfed2018,LHD2013,SPLENDID2019,Odyssey2017,semagrow2015,hasnain2012cataloguing,hasnain2014roadmap,sihjoin2011,aderis2011} achieve the goal of optimizing their query plan by first estimating the cardinality of the query's  triple patterns. Subsequently, they use this information to estimate the cardinality of the joins involved in the query. A cost model is then used to compute the cost of performing different query joins while considering network communication costs. One of the query plans with minimum execution costs is finally selected for result retrieval. Since the principal input for cost-based query planning is the cardinality estimates, the accuracy of these estimates is crucial to achieve a good query plan. 

The performance of federated SPARQL query processing engines has been evaluated in many recent studies \cite{fineGrainEval2014,Lusail2017,Odyssey2017,costfed2018,MULDER2017,semagrow2015,saleem2014hibiscus, daw2013, anapsid2011,umbrich2014, acosta2017diefficiency, biofed2017,iguana2017,safe1,safe2} using different federated benchmarks \cite{LargeRDFBench2018,FedBench2011,splodge2012,dbpeidabench2011,qfed2014,ExtLargeRDFBench2017, montoya2012benchmarking,Bizer09theberlin,schmidt2009sp}. Performance metrics, including query execution time, number of sources selected, source selection time, query planning time, continuous efficiency of query processing, answer completeness and correctness, time for the first answer, and throughput, are usually reported in these studies. While these metrics allow the evaluation of certain components (e.g., the source selection model), they cannot be used to evaluate the accuracy of the cardinality estimators of the cost-based federation engines. Consequently, they are unable to show how the estimated cardinality errors affect the overall query runtime performance of federation engines.


In this paper, we address the problem of measuring the accuracy of the cardinality estimators of federated SPARQL engines, as well as the effect of these errors on the overall query runtime performance. In particular, we propose metrics\footnote{Our proposed metric is open-source and available online at https://github.com/dice-group/CostBased-FedEval} for measuring errors in the cardinality estimations of (1) triple patterns, (2) joins between triple patterns, and (3) query plans. We correlate these errors with the overall query runtime performance of state-of-the-art cost-based SPARQL federation engines. The observed results show that these metrics are 
correlated with the overall runtime performances. In addition, we compare sate-of-the-art cost-based SPARQL federation engines using existing metrics pertaining to indexing, query processing, network, and overall query runtime using different evaluation setups. 

In summary, the contributions of this work are as follows:
\begin{itemize}
\item We propose metrics to measure the errors in cardinality estimations of cost-based federated engines. These metrics allow a fine-grained evaluation of cost-based federated SPARQL query engines and uncover novel insights about the performance of these types of federation engines that were not reported in previous works.
\item We measure the correlation between the values of the novel metrics and the overall query runtimes. We show that some of these metrics have a strong correlation with runtimes and can hence be used as predictors for the overall query execution performance. 
\item We present an empirical evaluation of five--- CostFed \cite{costfed2018}, Odyessey \cite{Odyssey2017}, SemaGrow \cite{semagrow2015}, LHD \cite{LHD2013} and SPLENDID \cite{SPLENDID2019}---state-of-the-art cost-based SPARQL federation engines on LargeRDFBench \cite{LargeRDFBench2018} by using the proposed metrics along with existing metrics, affecting the query runtime performance.
\end{itemize}
 
The rest of the paper is organized as follows: In Section \ref{RelatedWork}, we present related work. A motivating example is given in Section \ref{example}. In Section \ref{metrics}, we present our novel metrics for the evaluation of cost-based federation engines. In Section \ref{selection}, we give an overview of the cardinality estimators of selected cost-based federation engines.
The evaluation of these engines with proposed as well as existing metrics is shown in Section \ref{EvaluationandResults}. Finally, we conclude in Section \ref{sec:conclusion}.


\section{Related Work}
\label{RelatedWork}
In this section, we focus on the performance metrics used in the state-of-the-art to compare federated SPARQL query processing engines. 
Based on the previous federated SPARQL benchmarks \cite{LargeRDFBench2018,FedBench2011,splodge2012} and performance evaluations \cite{MULDER2017,Lusail2017,costfed2018,SPLENDID2019,semagrow2015,Odyssey2017,LHD2013,darq2008,anapsid2011,fedx2011, acosta2017diefficiency, biofed2017} (see Table \ref{tab:metric-stats} for an overview), the performance metrics used for federated SPARQL engines comparison can be categorized as: 

\newcolumntype{P}{>{\centering\arraybackslash}p{0.55cm}}
\begin{table*}[!htb]
\setlength\tabcolsep{1.55pt}
\centering
\footnotesize
\begin{tabular}{@{}llPPPPPPPPPPPPPPP@{}}
	\toprule &Engine& \multicolumn{2}{c}{Index} & \multicolumn{5}{c}{Processing} & \multicolumn{2}{c}{Network} & \multicolumn{2}{c}{Res}&
	\multicolumn{2}{c}{RS}&
	\multicolumn{2}{c}{Add}  \\  
	\cmidrule(lr){3-4} \cmidrule(lr){5-9} \cmidrule(lr){10-11} \cmidrule(lr){12-13} \cmidrule(lr){14-15} \cmidrule(lr){16-17} 
	&\texttt{} &	\texttt{Cr}	 &	\texttt{Gt}&	\texttt{Qp}& \texttt{\#Ts} &		\texttt{Qet}& \texttt{\#A}	  &		\texttt{Sst}	 &	\texttt{\#Tt}	 &	\texttt{\#Er}&		\texttt{Cu}&		\texttt{Mu}  & \texttt{Cp}  & \texttt{Ct} & \texttt{@T} & \texttt{@K} \\
	\midrule
	&CostFed \cite{costfed2018}  &  	\cmark & 	\cmark & 	 & 	\cmark & 	\cmark & 	\cmark & 	\cmark & 	 &   & 	 & 		 & \cmark &   & 	 & 	\\\midrule
& SPLENDID \cite{SPLENDID2019} &   	 & 	 &	 & 	\cmark & 		\cmark & 	 & 	 & 	 & 	\cmark &  & 	  & 	 &   & 	 	 & 	\\\midrule
& SemaGrow \cite{semagrow2015}	   &     & 	 & \cmark  & 	\cmark & 		\cmark & 	 & 	 &	 & 	 & 		 & 	  & 	
 &  &   	 & 	\\\midrule
& Odyssey \cite{Odyssey2017}  &    	 & 	&  	\cmark & 	\cmark & 		\cmark & 	 & 	 & 	\cmark & 	\cmark & & 	  & 	\cmark &  &		 & 	\\\midrule
& LHD \cite{LHD2013}	   &     & 	 & 	 & 	 & 		\cmark & 	 & 	 & 	\cmark & 	\cmark &	\cmark & 	\cmark &  \cmark &  &    	 & 	\\\midrule
& DARQ \cite{darq2008} & 	     & 	 & 	\cmark & 	 & 		\cmark & 	 & 	 & 	 & 	 & 	 & 	 &   &  &  	 & 	\\\midrule
& ANAPSID \cite{anapsid2011} &          & 	 & 	 	 & 	 & 		\cmark & 	 & 	 & 	 & 	 & 	 & 	 &
 &  &     & 	\\\midrule
 & HiBISCuS \cite{saleem2014hibiscus} & 		\cmark	      	 & 	\cmark	 &  & 		\cmark & 	\cmark & \cmark	 & 	\cmark	 & 	 &   &	 &    & \cmark&  	 	 & \\\midrule
& MULDER \cite{MULDER2017} &  	 & 	& 	 & 	 & 		\cmark & 	 & 	 & 	 & 	 &	 & 	  &  \cmark & \cmark &   	\cmark & 	\cmark\\\midrule
& FedX \cite{fedx2011} & 	 & 	& 	 & 	 & 		\cmark & 	\cmark & 	 & 	 & 	\cmark &	 & 	  &  &  &   	 & 	\\ \midrule
 & Lusail \cite{Lusail2017} & 	     & 	 & 	\cmark & 	 & 		\cmark & 	 & 	\cmark & 	 & 	 &   & 	 &   & &  	 	 & 	\\\midrule
  & BioFed \cite{biofed2017} & 	     & 	 & 	 & 	\cmark & 		\cmark & 	\cmark & 	\cmark & 	 & 	 &   & 	 &  \cmark &\cmark &  	 	 & 	\\\midrule
  &TopFed \cite{topfed2014}& 	      	 & 	 &  & 		\cmark & 	\cmark &  & 	 & 	 &   & 	 &   & &  	 	 & 	\\
  \midrule
  &SAFE \cite{safe1,safe2}& 	\cmark	      	 & 	\cmark	 &  & 		\cmark & 	\cmark & \cmark	 & 	\cmark	 & 	 &   & 	 &   & &  	 	 & 	\\

 \bottomrule
\end{tabular}
\caption{Metrics used in the existing federated SPARQL query processing systems, \textbf{Res:} Resource Related, \textbf{RS:} Result Set Related, \textbf{Add:} Additional, 
\textbf{Cr:} index compression ratio,
\textbf{Gt:} the index/summary generation time,
\textbf{Qp:} Query planning time,
\textbf{\#Ts:} total number of triple pattern-wise sources selected,
\textbf{Qet:} the average query execution time,
\textbf{\#A:} total number of SPARQL ASK requests submitted, \textbf{Sst:} the average source selection time, \textbf{\#Tt:} number of transferred tuples, \textbf{\#Er:} number of endpoint requests, \textbf{Cu:} CPU usage, \textbf{Mu:} Memory usage, \textbf{Cp:} Result Set completeness, \textbf{Ct:} Result Set correctness, \textbf{@T:} dief@t, \textbf{@K:} dief@k}
\label{tab:metric-stats}
\end{table*}

\begin{itemize}
   \item \textbf{Index-Related}: Index-assisted approaches \cite{fineGrainEval2014} make use of stored dataset statistics to generate an optimized query execution plan. The indexes are pre-computed by collecting information from available federated datasets. This is usually a one-time process. However, later updates are required to ensure the result-set completeness of the query processing. The index generation time and its compression ratio (w.r.t. overall dataset size) are important measures to be considered when devising index-assisted federated engines. 
    \item \textbf{Query-Processing-Related}: This category contains metrics related to the query processing capabilities of the federated SPARQL engines. The reported metrics in this category are the total number of triple-pattern-wise sources selected, number of \texttt{ASK} requests used to perform source selection, source selection time, query planning time, and overall query runtime. 
    \item \textbf{Network-Related:} Federated engines collect information from multiple data sources, e.g., SPARQL endpoints. Thus, it is important to minimize the network traffic generated by the engines during query processing. The number of transferred tuples and the number of endpoint requests generated by the federation engine are the two network-related metrics used in existing federated SPARQL query processing evaluations.  
    \item \textbf{Result-Set-Related:} Two systems are only comparable if they produce exactly the same results. Therefore, result set correctness and completeness are the two most important metrics in this category. 
    \item \textbf{Resource-Related:} The CPU and memory resources consumed during query processing dictate the query load an engine can bear. Hence, they are of importance when evaluating the performance of federated SPARQL engines.
     \item \textbf{Additional:} Two metrics dief@t and dief@k are proposed to measure continuous efficiency of query processing approaches.
    
\end{itemize}

All of these metrics are helpful to evaluate the performance of different components of federated query engines. However, none of these metrics can be used to evaluate the accuracy of the cardinality estimators of cost-based federation engines. Consequently,  studying the effect of  estimated cardinality errors on the overall query runtime performance of federation engines cannot be conducted based on these metrics. To overcome these limitations, we propose metrics for measuring errors in cardinality estimations of triple patterns, joins between triple patterns, and overall query plan, and show how these metrics are affecting the overall runtime performance of federation engines.


\def\redcolor{\color{red}}
\def\blackcolor{\color{black}}
\def\greencolor{\color{green}}
\def\bluecolor{\color{blue}}
\begin{figure*}[!htb]
\centering
\begin{subfigure}{0.18\textwidth}
  \centering
    \begin{lstlisting}[stepnumber=0,language=sparql,basicstyle=\small,escapechar=@]
SELECT * WHERE {
?s :p1 ?o1. 
@\aftergroup\redcolor@Cr(TP1):100@\aftergroup\greencolor@
Ce1(TP1):90 
@\aftergroup\bluecolor@Ce2(TP1):200@\aftergroup\blackcolor@

?s :p2 ?o2. 
@\aftergroup\redcolor@Cr(TP2):200@\aftergroup\greencolor@
Ce1(TP2):250 
@\aftergroup\bluecolor@Ce2(TP2):600@\aftergroup\blackcolor@

?s :p3 ?o3. 
@\aftergroup\redcolor@Cr(TP3):300@\aftergroup\greencolor@
Ce1(TP3):300
@\aftergroup\bluecolor@Ce2(TP3):500@\aftergroup\blackcolor@ }
    \end{lstlisting}
      \caption{Example query}
  \label{fig:query}
\end{subfigure}%
\begin{subfigure}{0.37\textwidth}
  \centering
  \includegraphics[scale=0.35]{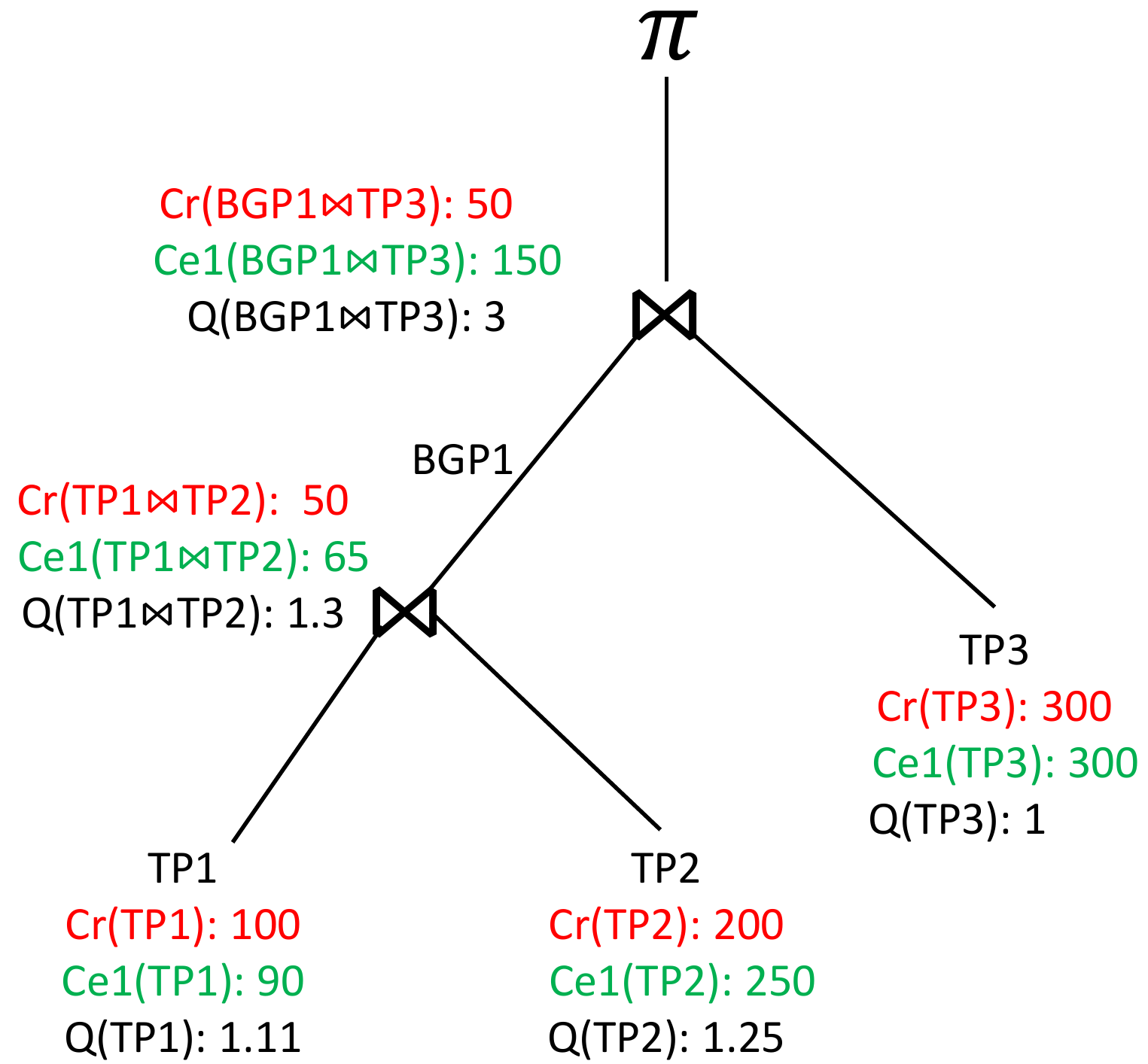}
  \caption{Engine 1 optimal query plan}
  \label{fig:e1p}
\end{subfigure}
\begin{subfigure}{0.37\textwidth}
  \centering
  \includegraphics[scale=0.35]{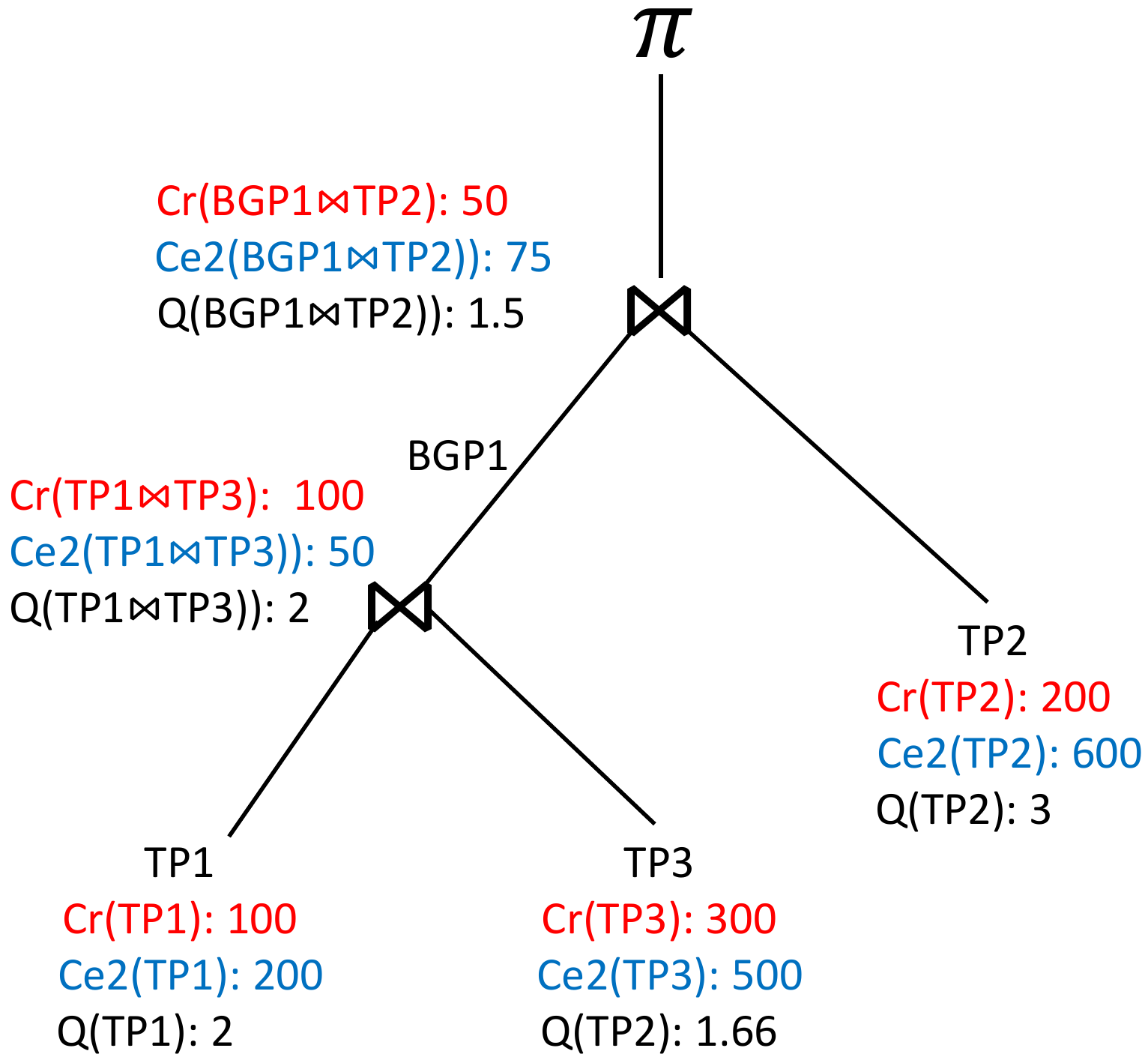}
  \caption{Engine 2 subOptimal query plan}
  \label{fig:e2p}
\end{subfigure}%
\caption{Motivating Example: A sample SPARQL query and the corresponding query plans of two different federation engines}
\label{fig:mexample}
\end{figure*}
\section{Motivating Example}\label{example}

In this section, we present an example to motivate our work and to understand the proposed metrics. We assume that the reader is familiar with the concepts of SPARQL and RDF, including the notions of a triple pattern, the joins between triple patterns, the cardinality (result size) of a triple pattern, and left-deep query execution plans. As aforementioned, most  cost-based SPARQL federation engines first estimate individual triple pattern cardinality and use this information to estimate the cardinality of joins found in the query. Finally, the query execution plan is generated by ordering the joins. In general, the optimizer first selects the triple patterns and joins with minimum estimated cardinalities \cite{costfed2018}.  

Figure \ref{fig:mexample} shows a motivating example containing a SPARQL query with three triple patterns---namely TP1, TP2 and TP3---and two joins. Consider two different cost-based federation engines with different cardinality estimators. Figure \ref{fig:query} shows the real (\texttt Cr) and estimated cardinalities (\texttt{Ce1} for Engine 1 and \texttt{Ce2} for Engine 2) for triple patterns of the query. Let us assume that both engines generate left-deep query plans by selecting triple patterns with the smallest cardinalities to perform their first join. The results of this join are then used to perform the second join with the remaining third triple pattern. By using actual cardinalities, the optimal query execution plan would be to first perform the join between TP1 and TP2 and then perform the second join with TP3. The same plan will be generated by Engine 1 as well, as 
shown in Figure \ref{fig:e1p}. The subOptimal plan generated by Engine 2 is given in Figure \ref{fig:e2p}. Note that Engine 2 did not select the optimal plan because of large errors in cardinality estimations of triple patterns and joins between triple patterns. 

The motivating example clearly shows that good cardinality estimations are essential to produce a better query plan.
The question we aim to answer pertains to how much the accuracy of cardinality estimations affects the overall query plan and the overall query runtime performance. To answer this question, the q-error (Q in Figure \ref{fig:mexample}) was introduced in \cite{qerror2009} in the database literature. In the next section, we define this measure and propose new metrics based on similarities to measure the overall triple patterns error $E_T$, overall joins error $E_J$ as well as overall query plan error $E_P$. 

\section{Cardinality Estimation-related Metrics}\label{metrics}
Now we formally define the q-error and our proposed metrics, namely $E_T$, $E_J$, $E_P$ to measure the overall error in cardinality estimations of triple patterns, joins between triple patterns and overall query plan error, respectively.  

\subsection{q-error}
\label{sec:qerror}
The q-error is the factor by which an estimated cardinality value differs
from the actual cardinality value \cite{qerror2009}. 

\begin{definition} [q-error] \label{def:Hypergraph}
Let $\vec{r} = (r_1,\dots,r_n) \in \mathbb{R}^{n} $ where $r_i > 0$ be a vector of real values and $\Vec{e} = (e_1,\dots,e_n) \in \mathbb{R}^{n}$ be the vector of the corresponding estimated values. By defining $\vec{e}/\vec{r} = \frac{\vec{e}}{\vec{r}} = (e_1/r_1,\ldots,e_n/r_n)$, then q-error of estimation e of r is given as 

\centering
{$||e/r||_Q = \max\limits_{1\leq i\leq n} ||e_i/r_i||_Q$}, where $||e_i/r_i||_Q = \max(e_i/r_i,r_i/e_i)$

\end{definition}
In this definition, over- and underestimations are treated symmetrically \cite{qerror2009}. In the motivating example given in Figure \ref{fig:mexample}, the real cardinality of TP1 is 100 (i.e., $Cr(TP1) = 100$) while the estimated cardinality by engine 1 for the same triple pattern is 90 (i.e., $Cr(TP1) = 90$). Thus, the q-error for this individual triple pattern is $\max(90/100,100/90) = 1.11$. The query's overall q-error of its triple patterns (see Figure \ref{fig:e1p}) is the maximum value of all the q-error values of triple patterns, i.e., $\max(1.11,1.25,1) = 1.25$. The q-error of the complete query plan would be the maximum q-error values in all triple patterns and joins used in the query plan, i.e., $\max(1.11, 1.25, 1, 1.3, 3) = 3$.  

The q-error makes use of the ratio instead of an absolute or quadratic difference and is hence able to capture the intuition that only relative differences matter for making planning decisions. In addition, the q-error provides a theoretical upper bound for the plan quality if the q-error of a
query is bounded. Since it only considers the maximum value amongst those calculated, it is possible that plans with good average estimations are regarded as poor by this measure. Consider the query plans given in Figure \ref{fig:e1p} and Figure \ref{fig:e2p}. Both have a q-error of 3, yet the query plan in Figure \ref{fig:e1p} is optimal, while the query plan in Figure \ref{fig:e2p} is not. To solve this problem, we introduce the additional metrics defined below. 

\subsection{Similarity Errors}
The overall similarity error of query triple patterns is formalised as follows: 
\begin{definition} [Triple Patterns Error \texttt{$E_T$} ]
Let Q be a SPARQL query containing triple patterns T = \{$TP_1,\ldots,TP_n$\}. Let $\vec{r} = (Cr(TP_1),\ldots,Cr(TP_n)) \in \mathbb{R}^{n}$ be the vector of real cardinalities of T and $\vec{e} = (Ce(TP_1), \ldots,Ce(TP_n)) \in \mathbb{R}^{n}$ be the vector of the corresponding estimated cardinalities of T. Then, we define our overall triple pattern error as follows:\\

$E_T$ = $  \frac{\lVert \vec{r} - \vec{e} \rVert} {\lVert \vec{r}\rVert + \lVert \vec{e} \rVert}= \frac{ \sqrt{\sum_{i=1}^{n} ( Cr(TP_i) - Ce(TP_i) )^2 }}{ \sqrt{\sum_{i=1}^{n}{(Cr(TP_i))^2}} + \sqrt{\sum_{i=1}^{n}{(Ce(TP_i))^2}} }$

 
\end{definition}

In the motivating example given in Figure \ref{fig:mexample}, the real cardinalities vector $\vec{r}$ = (100,200,300) and the Engine 1 estimated cardinalities vector $\vec{e}$ = (90,250,300). Thus, $E_T$ = $0.0658$. Similarly, the Engine 2 estimated cardinality vector is $\vec{e}$ = (200,500,600). Thus, Engine 2 achieves $E_T$ = $0.388$.   

\begin{definition} [Joins Error \texttt{$E_J$} ]
Let Q be a SPARQL query containing joins J = \{$J_1,\ldots,J_n$\}. Let $\vec{r} = (Cr(J_1),\ldots,Cr(J_n)) \in \mathbb{R}^{n}$ a vector of real cardinalities of J and $\vec{e} = (Ce(J_1),\ldots,Ce(J_n)) \in \mathbb{R}^{n}$ be the vector of the corresponding estimated cardinalities of J, then the overall joins error is defined by the same equation in Definition 2.


 
\end{definition}

\begin{definition} [Query Plan Error \texttt{$E_P$} ]
Let Q be a SPARQL query and TJ be the set of triple patterns and joins in Q. Let $\vec{r} = (r_1,\ldots,r_n) \in \mathbb{R}^{n} $ be a vector of real cardinalities of TJ and $\vec{e} = (e_1,\ldots,e_n) \in \mathbb{R}^{n} $ be the vector of corresponding estimated cardinalities of TJ, then the overall query plan error is defined by the same equation in Definition 2.
 \end{definition}
In the motivating example given in Figure \ref{fig:e1p}, the real cardinalities vector of all triple patterns and joins, $\vec{r}$ = (100,200,300,50,50) and the Engine 1 estimated cardinalities vectors $\vec{e}$ = (90,250,300,65,150). Thus, $E_P$ = 0.1391 for Engine 1. Engine 2
achieves $E_P$ = 0.3838. In these matrices, over- and underestimations are also treated symmetrically. The purpose of these definitions is to keep the lower bound at 0, which could be reached if $r = e$ (i.e., there is no error in the estimation), and the upper bound at 1, which could be reached if $e$ is much larger than $r$.

\section{Selected Federation Engines}
\label{selection}
In this section, we give a brief overview of the selected cost-based SPARQL federation engines. In particular, we describe how the cardinality estimations for triple patterns and joins between triple patterns are performed in these engines.  

\paragraph{\textbf{CostFed:}}
CostFed \cite{costfed2018} makes use of pre-computed statistics stored in index to estimate the cardinality of triple patterns and joins between triple patterns. CostFed benefits from both bind join ($\bowtie_b$) \cite{semagrow2015,fedx2011,SPLENDID2019} and symmetric hash join ($\bowtie_h$) \cite{anapsid2011} for joining the results of triple patterns. The decision of join selection is based on calculating the cost of both joins on query runtime. It creates three buckets for each distinct predicate used in the RDF dataset. These buckets are used for estimating the cardinality of query triple patterns. Furthermore, CostFed stores selectivity information that is used to estimate the cardinality of triple patterns as well as devising an efficient query plan. The CostFed query planner also considers the skew in the distribution of  objects and subjects across predicates. Separate cardinality estimation is used for Multi-valued predicates. Multi-valued predicates are the predicates that can have multiple values, as people can have multiple contact numbers or graduation schools. It performs a join-aware trie-based source selection, which uses common URI prefixes. 

Let  $D$ represent a dataset or source for short, $tp = <s, p, o>$ be a triple pattern having predicate $p$, and $R(tp)$ be the set of relevant sources for that triple pattern. The following notations are used to calculate the cardinality of tp.
\begin{itemize}
 \item $T(p,D)$ is the total number of triples with predicate $p$ in $D$.
 \item $\mathit{avgSS}(p,D)$ is the average subject selectivity of $p$ in $D$.
 \item $\mathit{avgOS}(p,D)$ is the average object selectivity of $p$ in $D$. 
 \item $tT(D)$ is the total number of triples in $D$. 
 \item $tS(D)$ is the total number of distinct subjects in $D$. 
 \item $tO(D)$ is the total number of distinct objects in $D$. 
\end{itemize}

From these notations the cardinality  $C(tp)$ of tp is calculated as follows (the predicate $\bound$ stands for bound):
\begin{flalign*}\small
\label{eq:card}
&\begin{cases}
\sum\limits_{\forall Di \in R(tp)} T(p, D_i) \times 1 &\\ \hspace{85pt} \Rightarrow \text{if \bound(p)$\wedge$!\bound(s) $\wedge$ !\bound(o),}\\
\sum\limits_{\forall Di \in R(tp)} T(p, D_i) \times avgSS(p,D_i) &\\ \hspace{85pt} \Rightarrow \text{if \bound(p) $\wedge$ \bound(s) $\wedge$ !\bound(o),}\\
\sum\limits_{\forall Di \in R(tp)} T(p, D_i) \times avgOS(p,D_i) &\\ \hspace{85pt} \Rightarrow \text{if \bound(p) $\wedge$ !\bound(s) $\wedge$ \bound(o),}\\
\sum\limits_{\forall Di \in R(tp)} tT(D_i) \times 1 &\\ \hspace{85pt} \Rightarrow \text{if !\bound(p) $\wedge$ !\bound(s) $\wedge$ !\bound(o),}\\
\sum\limits_{\forall Di \in R(tp)} tT(D_i) \times \frac{1}{tS(D_i)} &\\ \hspace{85pt} \Rightarrow \text{if !\bound(p) $\wedge$ \bound(s) $\wedge$ !\bound(o),}\\
\sum\limits_{\forall Di \in R(tp)} tT(D_i) \times \frac{1}{tO(D_i)} &\\ \hspace{85pt} \Rightarrow \text{if !\bound(p) $\wedge$ !\bound(s) $\wedge$ \bound(o),}\\
\sum\limits_{\forall Di \in R(tp)} tT(D_i) \times \frac{1}{tS(D_i) \times tO(D_i)} &\\ \hspace{85pt} \Rightarrow \text{if !\bound(p) $\wedge$ \bound(s) $\wedge$ \bound(o),}\\\\
1 \hspace{77pt} \Rightarrow \text{if \bound(p) $\wedge$ \bound(s) $\wedge$ \bound(o)}&
\end{cases}&
\end{flalign*}

A recursive definition is used to define the SPARQL expression $E$ \cite{semagrow2015,sparqlinfra2013} in the query planning phase and is defined as follows: all triple patterns are SPARQL expressions and if $E1$ and $E2$ are SPARQL expressions then $E1 \bowtie E2$ 
is also a SPARQL expression. The join cardinality of two expressions $E_1$ and $E_2$ is estimated as follows:
\begin{flalign*}
&C(E1 \bowtie E2)&\\ 
&\hspace{25pt} = M(E1) \times M(E2) \times Min(C(E1), C(E2))&
\end{flalign*}
where the average frequency of multi-valued predicates in the expression $E$ is defined as $M(E)$. In $M(E)$, $E$ is not the result of joins between triple patterns but the triple pattern itself. $M(E)$ is calculated using the following equation: 
\begin{flalign*}
&M(E) = &\\
&\begin{cases}
1/\sqrt{2} &\text{if \bound(p) $\wedge$ !\bound(s) $\wedge$ \bound(o),}\\
C(E)/distSbjs(p,\dss) &\text{if \bound(p) $\wedge$ !\bound(s) $\wedge$ !\bound(o) $\wedge$ j(s),}\\
C(E)/distObjs(p,\dss) &\text{if \bound(p)  $\wedge$ !\bound(o) $\wedge$ !\bound(s) $\wedge$ j(o),}\\
\text{1} &\text{other}
\end{cases}&
\end{flalign*}
If the subject of the triple pattern is involved in the join, it is defined as $j(s)$, and b(s), b(o), and b(p) are defined as bound subject, object, predicate respectively.  

\paragraph{\textbf{SPLENDID:}}
SPLENDID \cite{SPLENDID2019} also uses VoID statistics to generate a query execution plan. It uses a dynamic programming approach to produce a query execution plan. SPLENDID makes use of both hash ($\bowtie_h$) and bind ($\bowtie_b$) joins. 



Triple pattern cardinality is estimated as follows:
\begin{flalign*}
\operatorname{card}_{d}(?, p, ?) &=\operatorname{card}_{d}(p)  &\\ \operatorname{card}_{d}(s, ?, ?)&=|d| \cdot \operatorname{sel} . s_{d} &\\
\operatorname{card}_{d}(s, p, ?) &=\operatorname{card}_{d}(p) \cdot \operatorname{sel.s}_{d}(p)  &\\ 
\operatorname{card}_{d}(?, ?, o)&=|d| \cdot \operatorname{sel} . o_{d} &\\
\operatorname{card}_{d}(?, p, o) &=\operatorname{card}_{d}(p) \cdot \operatorname{sel} . o_{d}(p)  &\\
\operatorname{card}_{d}(s, ?, o)&=|d| \cdot \operatorname{sel} . s_{d} \cdot \operatorname{sel} . o_{d}
\end{flalign*}

where the $card_{d}(p)$ is the number of triple patterns in the data source $d$ having predicate $p$. The total number of triples in a data source $d$ is defined as $|d|$. If we have a bound predicate then the average selectivity of subject and object is defined as ${sel.s}_{d}(p)$ and ${sel.o}_{d}(p)$ respectively; if the predicate is not bound then the average selectivity of subject and object is defined as ${sel.s}_{d}$ and ${sel.o}_{d}$ respectively. 
In star-shaped queries, SPLENDID estimates the cardinality of triple patterns having the same subject separately. All triples with same subjects are grouped and then the minimum cardinality of all triple patterns with bound objects is calculated. Lastly, the cardinality of remaining triples with unbound objects is multiplied with the average selectivity of subjects and the minimum value. Formally, the equation is defined as:   
\begin{flalign*}
&\operatorname{card}_{d}(T)= &\\
&\hspace{10pt}\min \left(\operatorname{card}_{d}\left(T_{bound}\right)\right) \cdot \prod\left(sel.s_{d} \cdot \operatorname{card}_{d}\left(T_{unbound}\right)\right)&
\end{flalign*}
Join cardinality is estimated as follows: 
\begin{flalign*}
\operatorname{card}(q1 \bowtie q2) =  
&\operatorname{card}\left(q_{1}\right) \cdot \operatorname{card}\left(q_{2}\right) \cdot \operatorname{sel}_{\bowtie}(q 1, q 2)&
\end{flalign*} 

In these equations the ${sel}_{\bowtie}$ is the join selectivity of two input relations. It defines how many bindings are matched between two relations. 
SPLENDID uses the average selectivity of join variables as join selectivity.

\paragraph{\textbf{LHD:}}
LHD \cite{LHD2013} is a cardinality-based and index-assisted approach that aims to maximize parallel execution of sub-queries. It makes use of the VoID statistics for estimating the cardinality of triple patterns and joins between triple patterns. LHD only uses Bind joins for query execution. LHD implements a response-time-cost model by making an assumption that the response time of a query request is proportional to the total number of bindings transferred. LHD determines the total number of triples $t_d$, distinct subjects $s_d$ and objects $o_d$ from the VoID description of a dataset d. The VoID file also provides other information, such as the number of triples $t_{d.p}$, distinct subjects $s_{d.p}$ and distinct objects $o_{d.p}$ in the dataset d for a predicate p. The federation engine makes an assumption about uniform distribution of objects and subjects in datasets. Let's assume a triple pattern $T : \{S  P  O\}$ \footnote{In this section, the letters with a question mark (e.g. ?x)  denote a variable in an RDF triple, a literal value is represented by a lower-case letter (e.g. o) , and a variable or a literal value is defined by an upper-case letter (e.g. S)}, the function to get the set of relevant datasets of T is defined as $S(T)$, the selectivity of x with respect to $S(T)$ is defined as $selT(x)$, and the cardinality of x with respect to $S(T)$ is defined as $cardT(x)$.

For single triple pattern cardinality estimation, the selectivity of each part is estimated as follows:
\begin{flalign*}
&\operatorname{sel}_{T}(S) =  &\\
&\hspace{25pt}\left\{\begin{array}{ll}
\frac{\sum_{d \in S(T)} t_{d} / s_{d}}{\sum_{d \in S(T)} s_{d}} & \text { if } \operatorname{var}(P) \wedge \neg \operatorname{var}(S) \\
\frac{\sum_{d \in S(T)} t_{d \cdot p}/s_{d \cdot p}}{\sum_{d \in S(T)} s_{d \cdot p}} & \text { if } P=p \wedge \neg \operatorname{var}(S) \\
1 & \text { if } \operatorname{var}(S)
\end{array}\right.&\\
&\operatorname{sel}_{T}(P) = &\\
&\hspace{25pt}\left\{\begin{array}{ll}
\frac{\sum_{d \in S(T)} t_{d . p}}{\sum_{d \in S(T)} t_{d}} & \text { if } P=p \\
1 & \text { if } \operatorname{var}(P)
\end{array}\right.&\\
&\operatorname{sel}_{T}(O) = &\\
&\hspace{25pt}\left\{\begin{array}{ll}
\frac{\sum_{d \in S(T)} t_{d} / o_{d}}{\sum_{d \in S(T)} o_{d}} & \text { if } \operatorname{var}(P) \wedge \neg 
\operatorname{var}(O) \\
\frac{\sum_{d \in S(T)} t_{d . p} / o_{d . p}}{\sum_{d \in S(T)} o_{d . p}} & \text { if } P=p \wedge \neg \operatorname{var}(O) \\
1& \text { if } \operatorname{var}(O)
\end{array}\right.&
\end{flalign*}
After calculating the selectivity of each part, LHD estimates the cardinality of the triple pattern as follows:
\begin{flalign*}
&\operatorname{card}(T)=t \cdot \operatorname{sel}_{T}(S) \cdot \operatorname{sel}_{T}(P) \cdot \operatorname{sel}_{T}(O)&
\end{flalign*}

Given two triple patterns T1 and T2, LHD calculates the join selectivity by using the following equations:
\begin{flalign*}
&\operatorname{sel}\left(T_{1} \bowtie T_{2}\right)=&\\
&\hspace{15pt}\left\{\begin{array}{ll}
\frac{\sum_{d \in S\left(T_{1}\right)} s_{d . p1} \cdot \sum_{d \in S\left(T_{2}\right)} s_{d . p2}}{\sum_{d \in S\left(T_{1}\right)}{s_{d} \cdot \sum_{d \in S\left(T_{2}\right)} s_{d}}} & \text { if joined on } S_{1}=S_{2}\\ \\
\frac{\sum_{d \in S\left(T_{1}\right)} o_{d . p1} \cdot \sum_{d \in S\left(T_{2}\right)} o_{d . p2}}{\sum_{d \in S\left(T_{1}\right)}{o_{d} \cdot \sum_{d \in S\left(T_{2}\right)} o_{d}}} & \text { if joined on } O_{1}=O_{2}\\ \\
\frac{\sum_{d \in S\left(T_{1}\right)} s_{d . p1} \cdot \sum_{d \in S\left(T_{2}\right)} o_{d . p2}}{\sum_{d \in S\left(T_{1}\right)}{s_{d} \cdot \sum_{d \in S\left(T_{2}\right)} o_{d}}} & \text { if joined on } S_{1}=O_{2}\\ \\
1 & \text { if no shared variables. }
\end{array}\right.&
\end{flalign*}

Using the join selectivity values, join cardinality is estimated by the following equation:
\begin{flalign*}
&\operatorname{card}\left(T_{1} \bowtie T_{2} \bowtie \ldots \bowtie T_{n}\right) &\\
&\hspace{35pt}=\prod_{i=1}^{n} \operatorname{card}\left(T_{i}\right) \cdot \operatorname{sel}\left(T_{1} \bowtie T_{2} \bowtie \ldots \bowtie T_{n}\right) &
\end{flalign*}

\paragraph{\textbf{SemaGrow:}}
SemaGrow \cite{semagrow2015} query planning is based on VoID\footnote{VoID vocabulary: \url{https://www.w3.org/TR/void/}} statistics \cite{void2009} about datasets. It makes use of the VoID index as well as SPARQL ASK queries to perform source selection. Three types of joins, i.e, bind, merge and hash, are used during the query planning. The selection to perform the required join operation is based on a cost function. It uses a reactive model for retrieving results of the joins as well as individual triple patterns. As with CostFed, SemaGrow recursively defines SPARQL expressions. Given a data source S, the cardinality estimations of triple patterns and joins are explained below.

SemaGrow contains a Resource discovery component, which returns the list of relevant sources to a triple pattern along with statistics. The statistics related to the data source include (1) the number of estimated distinct subjects, predicates and objects matching the triple pattern, and (2) the number of triple patterns in the data sources matching the triple pattern. The cardinality	of a triple	pattern	is provided	by the Resource Discovery component. On the other hand, for more complex expressions, SemaGrow needs to make an estimation based on	available statistics. In order to estimate complex expressions based on the aforementioned basic statistics, SemaGrow	adopts the formulas described by LHD \cite{LHD2013}. The cardinality of each expression (E) in a data source S, is defined as ${Card}([E], S)$).

For estimating the join cardinality we need to calculate the join selectivity ($\text { JoinSel }([E1] \bowtie [E2])$), which is given as follows:
\begin{flalign*}
&\text { JoinSel }\left([E1] \bowtie [E2]\right) =&\\ 
&\hspace{55pt} \min \left(\text { JoinSel }[E1], \text { JoinSel }[E2]\right)&\\
&\text { JoinSel }([T]) = \min \left({1}/{d_{i}}\right) &
\end{flalign*}
In these equations, E1 and E2 reside any join expressions or triple patterns. The T is a single triple pattern. $d_{i}$ represents the number of distinct values for the i-st join attribute in a T.  Hence, the join cardinality is given as follows:
\begin{flalign*}
&\operatorname{Card}([E1] \bowtie [E2], S)=&\\
&\operatorname{Card}([E1], S) \cdot \operatorname{Card}([E2], S) \cdot \text{ JoinSel } ([E1] \bowtie [E2])&
\end{flalign*}

\paragraph{\textbf{Odyssey:}}
Odyssey \cite{Odyssey2017} 
makes use of the distributed characteristic sets (CS) \cite{neumann2011characteristic} and characteristic pair (CP) \cite{gubichev2014exploiting} statistics to estimate cardinalities. Odyssey estimates the cardinality of each type of query differently using these statistics. 
For star-shaped queries, where the subject (or object) is the same for all joining triple patterns, estimated cardinality for a given set of properties P (predicates of joining triple patterns) is computed using CSs $C_{j}$ containing all these properties. The common subject (or object) is defined as an entity.  CSs can be computed by scanning once a dataset’s triples are sorted by subject; after all the entity properties have been scanned, the entity’s CS is identified. For each CS C, Odyssey computes statistics, i.e., $(count(C))$ represents the number of entities sharing C and $(occurrences(p, C))$ represents the number of triples with predicate p occurring with these entities.

Odyssey represents $\text{estimatedCardinality}_{Distinct}(P)$ as the estimated cardinality of queries that contain distinct keywords, and $\text{estimatedCardinality}(P)$ as the estimated cardinality of those queries that do not contain the distinct keyword. Formally, estimated cardinality for star-shaped queries is defined as follows:
\begin{flalign*}\small
&\text { estimatedCardinality }_{Distinct}(P)=\sum_{P \subseteq C_{j}}\left(\operatorname{count}\left(C_{j}\right)\right)&\\
&\text { estimatedCardinality }(P)=&\\
&\hspace{25pt}\sum_{P \subseteq C_{j}}\left(\operatorname{count}\left(C_{j}\right) \cdot \prod_{p_{i} \in P} \frac{\text { ocurrences }\left(p_{i}, C_{j}\right)}{\operatorname{count}\left(C_{j}\right)}\right)&
\end{flalign*}


For arbitrary-shaped queries, Odyssey also considers the connections (links) between different CSs. Characteristic pairs (CPs) help in describing the links between Characteristic sets (CSs) using properties. For entities $e1$ and $e2$ the link is defined as $(CSs(e1), CSs(e2), p)$, given that $(e1, p,e2) \in s$, where s is data source. 
The number of links between two $CSs$: $C_{i}$ and $C_{j}$, through a property p is represented in statistics, which is defined as: – $count(Ci,Cj,p)$. The equation for estimating the cardinality (pairs of entities with a set of properties $P_{k}$ and $P_{l}$) for complex-shaped queries is defined as:
\begin{flalign*}
&\begin{array}{l} 
\text {estimatedCardinality}\left(P_{k}, P_{l}, p\right)=\\\\
\sum_{P_{k} \subseteq C_{i} \wedge P_{l} \subseteq C_{j}}(\operatorname{count}\left(C_{i}, C_{j}, p\right) 
\cdot \prod_{p_{k} \in P_{k}-\{p\}}\\\\
\hspace{35pt}\left(\frac{\text { ocurrences }\left(p_{k}, C_{i}\right)}{\operatorname{count}\left(C_{i}\right)}\right) \cdot \prod_{p_{l} \in P_{l}}\left(\frac{\text { ocurrences }\left(p_{l}, C_{j}\right)}{\operatorname{count}\left(C_{j}\right)}\right))
\end{array}&
\end{flalign*}
In order to reduce the complexity, Odyssey treats each star-shaped query as a single meta-node; assuming that the order of joins has already optimized within the star-shaped sub-queries. It uses Characteristics Pairs (CPs) to estimate the cardinality of joins between star-shaped queries (meta-nodes) and uses dynamic programming (DP) to optimize the join order and find the optimal plan. 
\section{Evaluation and Results}\label{EvaluationandResults}
In this section, we discuss the results we obtained in our evaluation. All results are also available at the project homepage. First, we evaluate our novel metrics in terms of how they are correlated with the overall query runtime performance of state-of-the-art federated query engines. Thereafter, we compare existing cost-based SPARQL federation engines using the proposed metrics and discuss the evaluation results. 

\subsection{Experiment Setup and Hardware:}
\paragraph{\textbf{Benchmarks Used:}}
In our experiments, we used the state-of-the-art benchmark for federated engines dubbed LargeRDFBench~\cite{LargeRDFBench2018}.  
LargeRDFBench comprises a total of 40 queries (including all queries from  FedBench~\cite{FedBench2011}): 14 simple queries (S1-S14) from FedBench, 10 complex queries (C1-C10), 8 complex and high-sources queries (CH1-CH8), and 10 large data queries (L1-L10). Simple queries are
 fast to execute and include the smallest number of triple patterns, which ranges from 2 to 7 \cite{LargeRDFBench2018}. Complex queries are more challenging and take more time to execute compared to simple queries \cite{LargeRDFBench2018}. The queries in this category have at least 8 triple patterns and contain more joins and SPARQL operators than  simple queries. The complex and high-sources queries are even more challenging as they need to retrieve results from more data sources and they have more triple patterns, joins and SPARQL operators than the  simple and complex queries \cite{LargeRDFBench2018}. 

We used all queries except the large data queries (L1-L10) in our experiments. The reason for not using  L1-L10 was that the evaluation results presented in \cite{LargeRDFBench2018} show that most engines are not yet able to execute these queries.  LargeRDFBench comprises of 13 real-world RDF datasets of varying sizes. We loaded each dataset into a Virtuoso 7.2 server.  

\paragraph{\textbf{Cost-based Federation Engines:}}We evaluated five--- CostFed \cite{costfed2018}, Odyessey \cite{Odyssey2017}, SemaGrow \cite{semagrow2015}, LHD \cite{LHD2013} and SPLENDID \cite{SPLENDID2019}---state-of-the-art cost-based SPARQL federation engines. To the best of our knowledge, these are most of the currently available, open-source cost-based federation engines.

\paragraph{\textbf{Hardware Used:}} Each Virtuoso was deployed on a physical machine (32 GB RAM, Core i7 processor and 500 GB hard disk). We ran the selected federation engines on a local client machine with the same specification. Our experiments were run in a local environment where the network cost is negligible. This is the standard setting used in the original LargeRDFBench. Note that the accuracy of the cardinality estimators of the federated SPARQL query processing is independent of the network cost.

\paragraph{\textbf{Warm-up and Number of Runs:}}
In each experiment, we warmed up each federation engine for 10 minutes by executing the Linked Data (LD1-LD10) queries from FedBench. Experiments were run three times and the results were averaged. The query timeout was set to 30 minutes. 

\paragraph{\textbf{Metrics:}} We present results for the (1) q-error of triple patterns, (2) q-error of joins between triple patterns, (3) q-error of overall query plans, (4) errors of triple patterns, (5) errors of joins between triple patterns, (6) errors of overall query plans, (7) overall query runtimes, (8) number of tuples transferred (intermediate results), (9) source selection related metrics, and (10) quality of plans generated by query planner of each engine. In addition, we used Spearman's correlation coefficient to measure the correlation between the proposed metrics and the overall query runtimes. The Spearman test is designed to assess how well the dependency between two variables can be described using a monotonic function. While the Pearson test could also be used, we preferred the Spearman test because it is parameter-free and tests at rank level. 
We used simple linear and robust regression models to compute the correlation. 

\definecolor{ne}{HTML}{a50026}
\definecolor{nd}{HTML}{d73027}
\definecolor{nc}{HTML}{f46d43}
\definecolor{nb}{HTML}{fdae61}
\definecolor{na}{HTML}{fee08b}
\definecolor{no}{HTML}{ffffbf}
\definecolor{pa}{HTML}{d9ef8b}
\definecolor{pb}{HTML}{a6d96a}
\definecolor{pc}{HTML}{66bd63}
\definecolor{pd}{HTML}{1a9850}
\definecolor{pe}{HTML}{006837}

\newcommand{\corrpa}[1]{\cellcolor{pa}\textcolor{black}{$#1$}}
\newcommand{\corrpb}[1]{\cellcolor{pb}\textcolor{black}{$#1$}}
\newcommand{\corrpc}[1]{\cellcolor{pc}\textcolor{black}{$#1$}}
\newcommand{\corrpd}[1]{\cellcolor{pd}\textcolor{white}{$#1$}}
\newcommand{\corrpe}[1]{\cellcolor{pe}\textcolor{white}{$#1$}}
\newcommand{\corrno}[1]{\cellcolor{no}\textcolor{black}{$#1$}}
\newcommand{\corrna}[1]{\cellcolor{na}\textcolor{black}{$#1$}}
\newcommand{\corrnb}[1]{\cellcolor{nb}\textcolor{black}{$#1$}}
\newcommand{\corrnc}[1]{\cellcolor{nc}\textcolor{black}{$#1$}}
\newcommand{\corrnd}[1]{\cellcolor{nd}\textcolor{white}{$#1$}}
\newcommand{\corrne}[1]{\cellcolor{ne}\textcolor{white}{$#1$}}

\subsection{Regression Experiments}
\begin{figure*}[!htb]
\centering
\includegraphics[width=0.98\linewidth,height=20cm]{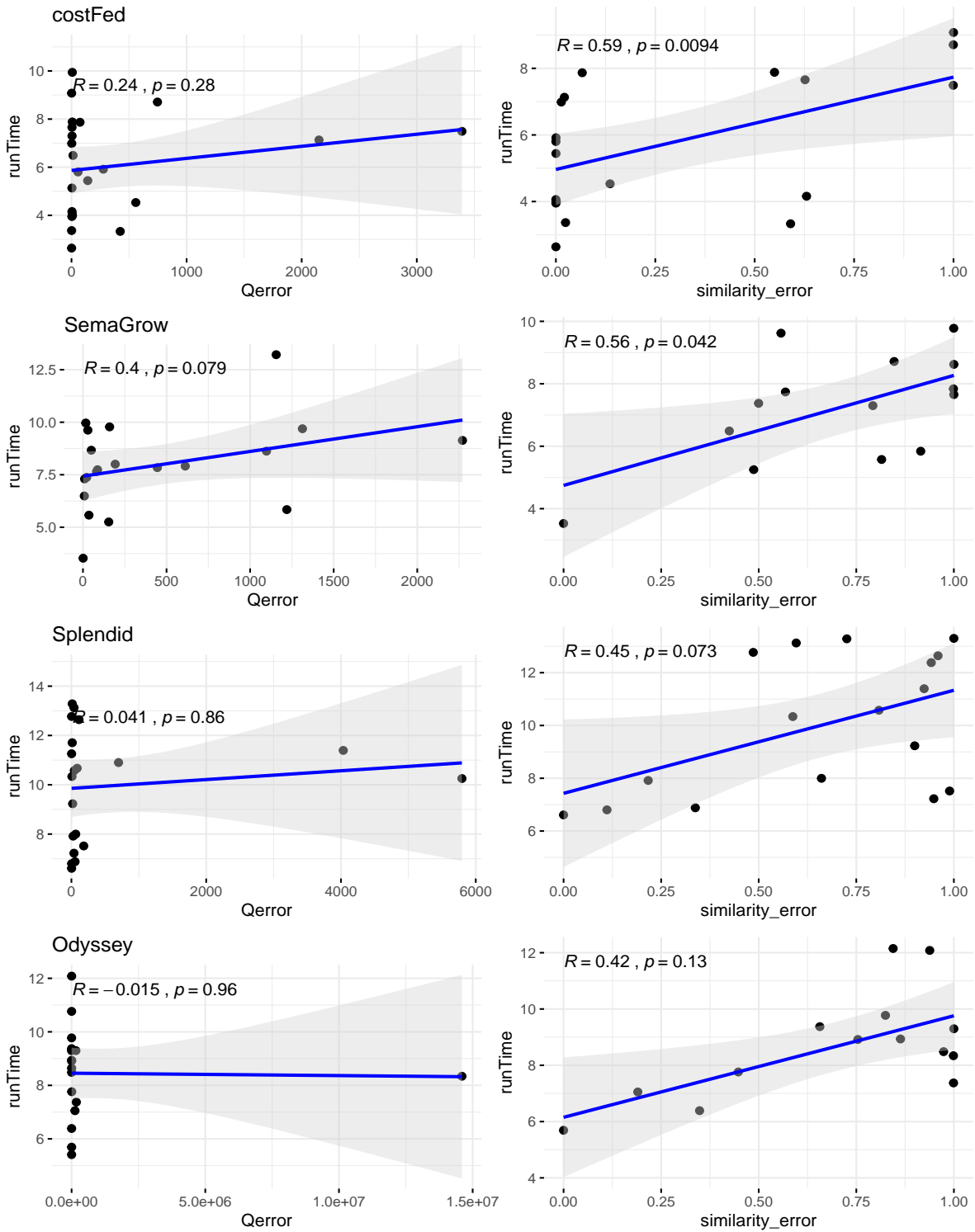}
\caption{q-error and similarity error vs. runtime (Simple Linear Regression Analysis). The grey shaded areas represent the confidence intervals (bands) in regression line.}
\label{fig:mvsr1}
\end{figure*}

\begin{figure*}[!htb]
\centering
\includegraphics[width=0.98\linewidth,height=20cm]{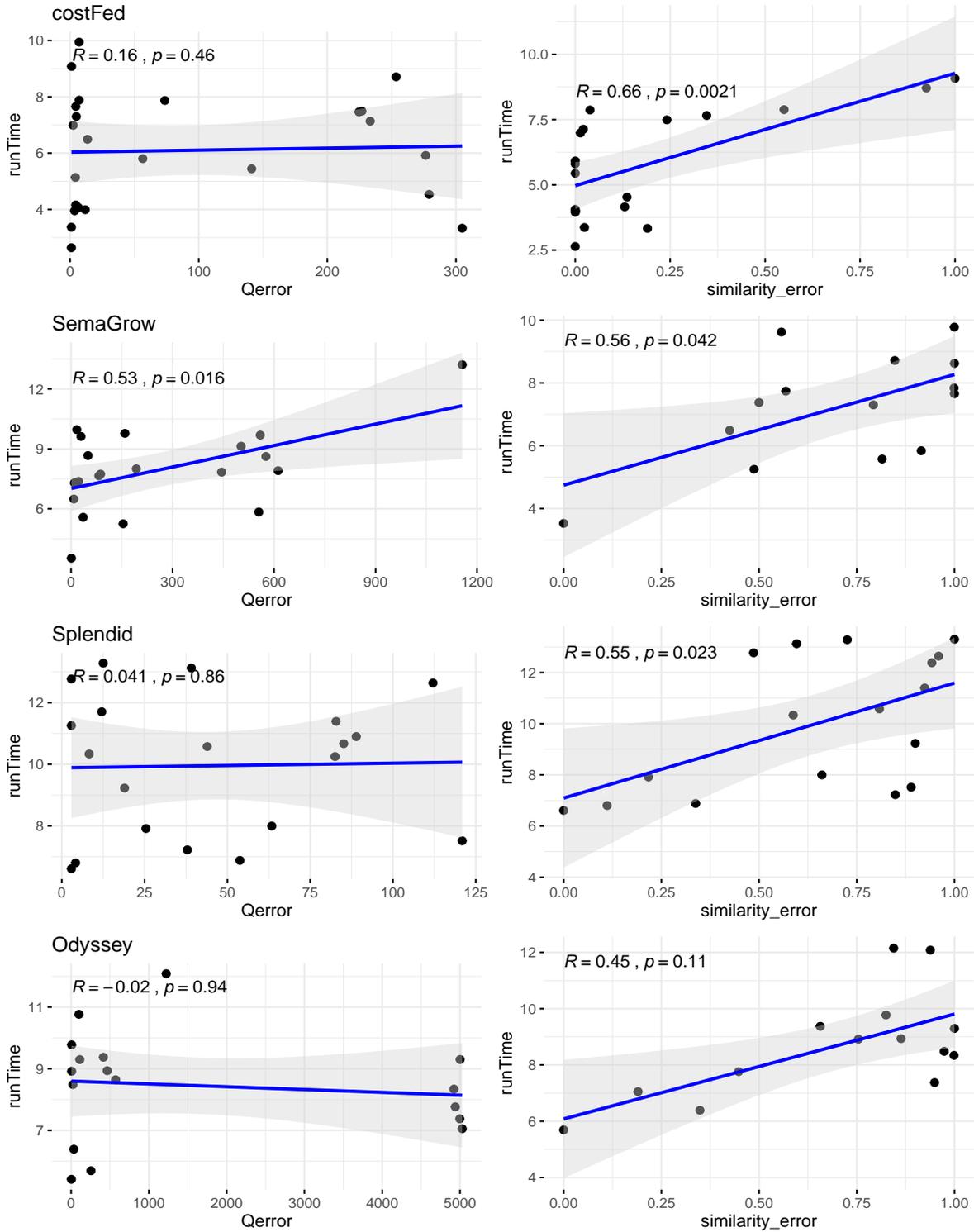}
\caption{q-error and similarity error vs. runtime (Robust Regression Analysis). The grey areas represent the confidence intervals (bands) in regression line. }
\label{fig:mvsr2}
\end{figure*}

Throughout our regression experiments, our null hypothesis was that there is no correlation between the runtimes of queries and error measurements (i.e., q-error or similarity error) used in the experiments. We began by investigating the dependency between the metrics we proposed and the overall query runtime performance of the federation engines selected for our experiments. Figure \ref{fig:mvsr1} shows the results of a simple linear regression experiment aiming to compute the dependency between the q-error and similarity errors and the overall query runtimes. For a particular engine, the left figure shows the dependency between the q-error and overall runtime while the right figure in the same row shows the result of the correlation of runtime with  similarity error. The higher coefficients (dubbed $R$ in the figure) computed in the experiments with similarity errors suggest that it is likely that the similarity errors are a better predictor for runtime. The positive value of the coefficient suggests that an increase in similarity error also means an increase in the overall runtime.  
It can be observed from the figure that outliers are potentially contaminating the results. Hence, we applied robust regression \cite{IRLS1977,IRLS1990,rousseeuw2005robust} using the Huber loss function \cite{Huber1992} in a second series of experiments to lessen the effect of the outliers on the results (especially for q-errors) (see Figure \ref{fig:mvsr2}). We observe that after removing outliers using robust regression, the average $R$-values of the  similarity-based error correlation further increases. 
The lower p-values in the similarity-error-based experiments further confirm that our metrics are more likely to be a better predictor for runtime than the q-error. The reason for this result is that our measure exploits more information and is hence less affected by outliers. This is not the case for the q-error, which can be perturbed significantly by a single outlier.


\begin{table*}[!htb]
\centering
\footnotesize
\setlength\tabcolsep{5.5pt}
\begin{tabular}{@{}clrrrrrrrrrrrrr@{}}
  \toprule
  
  & Rank &     
  \multicolumn{1}{c}{1} &     
  \multicolumn{1}{c}{2} &     
  \multicolumn{1}{c}{3} &
  \multicolumn{1}{c}{}  &
  \multicolumn{1}{c}{4} &     
  \multicolumn{1}{c}{5} &     
  \multicolumn{1}{c}{6}&     
  \multicolumn{1}{c}{}
  \\&&\multicolumn{4}{c}{Similarity Error}&\multicolumn{4}{c}{q-error}\\
  	\cmidrule(lr){3-6} \cmidrule(lr){7-10} 
  & Feature &
  \multicolumn{1}{c}{$E_J$} &      
  \multicolumn{1}{c}{$E_P$} &     
  \multicolumn{1}{c}{$E_T$} &
  \multicolumn{1}{c}{Average} &
  \multicolumn{1}{c}{$Q_J$} &     
  \multicolumn{1}{c}{$Q_P$} &     
  \multicolumn{1}{c}{$Q_T$}  &     
  \multicolumn{1}{c}{Average}  \\ \midrule
	\parbox[t]{2.8mm} {\multirow{4}{*}{\rotatebox[origin=c]{90}{\textbf{F.Q Engines}}}}&

	  CostFed      &  
	  \corrpb{0.23} &  \corrpc{0.59} &  \corrpc{0.43} &\corrpc{0.42} & \corrpa{0.14}  &  \corrpb{0.26}  & \corrpa{0.1} & \corrpa{0.17}  \\
	& SemaGrow    &  
	\corrpb{0.33} &  \corrpb{0.33} &  \corrpb{0.33}  & \corrpb{0.33} &   \corrpc{0.47} &  \corrpb{0.37} & \corrpa{0.001} & \corrpb{0.28} \\
	& ODYSSEY          & 
	\corrpa{0.11} & \corrpa{0.14} & \corrpc{0.55} & \corrpb{0.26} &  \corrpa{0.01} &  \corrpa{0.03} & \corrna{-0.06} & \corrna{-0.01}\\
	& SPLENDID      &  
	 \corrpb{0.3} &  \corrpc{0.4} &  \corrpb{0.24} & \corrpb{0.32}  &    \corrpa{0.17} &  \corrpa{0.1} & \corrpb{0.24} & \corrpa{0.17}\\
	& LHD        &  
	\corrpa{0.16} &  \corrpb{0.28} &  \corrna{-0.2}  & \corrpa{0.08} &   \corrpc{0.51} & \corrpa{0.11} & \corrpa{0.04} & \corrpb{0.22} \\ \midrule
	& Average       &  
	\corrpb{0.22} &  \corrpb{0.35} &  \corrpb{0.27}  &  \corrpb{0.28} &  \corrpb{0.26} & \corrpa{0.17} &  \corrpa{0.06} & \corrpa{0.17}\\ \bottomrule
\end{tabular}
\caption{Spearman's rank correlation coefficients between query plan features and query runtimes for all queries.}
\label{tab:correlation}
\end{table*}


\begin{table*}[!htb]
\centering
\footnotesize
\setlength\tabcolsep{5.5pt}
\begin{tabular}{@{}clrrrrrrrrrrrrr@{}}
  \toprule
  
  & Rank &     
  \multicolumn{1}{c}{1} &     
  \multicolumn{1}{c}{2} &     
  \multicolumn{1}{c}{3} &
  \multicolumn{1}{c}{}  &
  \multicolumn{1}{c}{4} &     
  \multicolumn{1}{c}{5} &     
  \multicolumn{1}{c}{6}&     
  \multicolumn{1}{c}{}
  \\&&\multicolumn{4}{c}{Similarity Error}&\multicolumn{4}{c}{q-error}\\
  	\cmidrule(lr){3-6} \cmidrule(lr){7-10} 
  & Feature &
  \multicolumn{1}{c}{$E_J$} &      
  \multicolumn{1}{c}{$E_P$} &     
  \multicolumn{1}{c}{$E_T$} &
  \multicolumn{1}{c}{Average} &
  \multicolumn{1}{c}{$Q_J$} &     
  \multicolumn{1}{c}{$Q_P$} &     
  \multicolumn{1}{c}{$Q_T$}  &     
  \multicolumn{1}{c}{Average}  \\ \midrule
	\parbox[t]{2.8mm} {\multirow{4}{*}{\rotatebox[origin=c]{90}{\textbf{F.Q Engines}}}}&

	  CostFed      &  
	  \corrpb{0.54} &  \corrpd{0.61} &  \corrpb{0.36} &\corrpc{0.5} & \corrpa{0.11}  &  \corrpb{0.23}  & \corrpa{0.05} & \corrpa{0.13}  \\
	& SemaGrow    &  
	\corrpc{0.44} &  \corrpc{0.56} &  \corrpc{0.43}  & \corrpc{0.48} &   \corrpc{0.49} &  \corrpc{0.40} & \corrna{-0.02} & \corrpb{0.29} \\
	& ODYSSEY          & 
	\corrpb{0.22} & \corrpc{0.42} & \corrpc{0.53} & \corrpb{0.39} &  \corrna{-0.04} &  \corrna{-0.01} & \corrnb{-0.20} & \corrna{-0.08}\\
	& SPLENDID      &  
	 \corrpb{0.35} &  \corrpc{0.45} &  \corrpb{0.27} & \corrpb{0.36}  &    \corrpa{0.12} &  \corrpa{0.04} & \corrpb{0.21} & \corrpa{0.12}\\\midrule
	& Average       &  
	\corrpb{0.39} &  \corrpc{0.51} &  \corrpc{0.40}  &  \corrpc{0.43} &  \corrpa{0.17} & \corrpa{0.17} &  \corrpa{0.01} & \corrpa{0.12}\\ \bottomrule
\end{tabular}
\caption{Spearman's rank correlation coefficients between query plan features and query runtimes after linear regression (only for common queries between all systems).}
\label{tab:c_correlation}
\end{table*}

\begin{table*}[!htb]
\centering
\footnotesize
\setlength\tabcolsep{5.5pt}
\begin{tabular}{@{}clrrrrrrrrrrrrr@{}}
  \toprule
  
  & Rank &     
  \multicolumn{1}{c}{1} &     
  \multicolumn{1}{c}{2} &     
  \multicolumn{1}{c}{3} &
  \multicolumn{1}{c}{}  &
  \multicolumn{1}{c}{4} &     
  \multicolumn{1}{c}{5} &     
  \multicolumn{1}{c}{6}&     
  \multicolumn{1}{c}{}
  \\&&\multicolumn{4}{c}{Similarity Error}&\multicolumn{4}{c}{q-error}\\
  	\cmidrule(lr){3-6} \cmidrule(lr){7-10} 
  & Feature &
  \multicolumn{1}{c}{$E_J$} &      
  \multicolumn{1}{c}{$E_P$} &     
  \multicolumn{1}{c}{$E_T$} &
  \multicolumn{1}{c}{Average} &
  \multicolumn{1}{c}{$Q_J$} &     
  \multicolumn{1}{c}{$Q_P$} &     
  \multicolumn{1}{c}{$Q_T$}  &     
  \multicolumn{1}{c}{Average}  \\ \midrule
	\parbox[t]{2.8mm} {\multirow{4}{*}{\rotatebox[origin=c]{90}{\textbf{F.Q Engines}}}}&

	  CostFed      &  
	  \corrpd{0.60} &  \corrpd{0.66} &  \corrpd{0.62} &\corrpd{0.63} & \corrpa{0.16}  &  \corrpa{0.16}  & \corrpa{0.16} & \corrpa{0.16}  \\
	& SemaGrow    &  
	\corrpc{0.56} &  \corrpc{0.56} &  \corrpc{0.57}  & \corrpc{0.56} &   \corrpd{0.60} &  \corrpc{0.53} & \corrpc{0.57} & \corrpc{0.56} \\
	& ODYSSEY          & 
    \corrpb{0.25} &  \corrpc{0.45} &  \corrpc{0.59} & \corrpc{0.43}  &    \corrna{-0.04} &  \corrna{-0.02} & \corrnb{-0.20} & \corrna{-0.08}\\

	& SPLENDID      &  
 	\corrpc{0.49} & \corrpc{0.55} & \corrpb{0.20} & \corrpb{0.38} &  \corrpa{0.14} &  \corrpa{0.041} & \corrpa{0.18} & \corrpa{0.12}\\ \midrule
	& Average       &  
	\corrpc{0.45} &  \corrpc{0.56} &  \corrpc{0.49}  &  \corrpc{0.50} &  \corrpb{0.22} & \corrpa{0.18} &  \corrpa{0.18} & \corrpa{0.19}\\ \bottomrule
\end{tabular}
\caption{Spearman's rank correlation coefficients between query plan features and query runtimes after robust regression (only for common queries between all systems).
$E_J$: Similarity Error of Joins,
$E_P$: Similarity Error of overall query plan,
$E_T$: Similarity Error of Triple Patterns,
$Q_J$: q-error of Joins,
$Q_P$: q-error of overall query plan,
$Q_T$: q-error Error of Triple Patterns,
F.Q: Federated Query.
Correlations and colors ($-+$):
$0.00 \mydots 0.19$ very weak    (\textcolor{na}{$\CIRCLE$}\textcolor{pa}{$\CIRCLE$}),
$0.20 \mydots 0.39$ weak         (\textcolor{nb}{$\CIRCLE$}\textcolor{pb}{$\CIRCLE$}),
$0.40 \mydots 0.59$ moderate     (\textcolor{nc}{$\CIRCLE$}\textcolor{pc}{$\CIRCLE$}),
$0.60 \mydots 0.79$ strong       (\textcolor{nd}{$\CIRCLE$}\textcolor{pd}{$\CIRCLE$}), 
$0.80 \mydots 1.00$ very strong  (\textcolor{ne}{$\CIRCLE$}\textcolor{pe}{$\CIRCLE$}).}
\label{tab:cr_correlation}
\end{table*}

\begin{figure*}[!htb]
\centering
\begin{subfigure}{.48\textwidth}
  \includegraphics[scale=0.48]{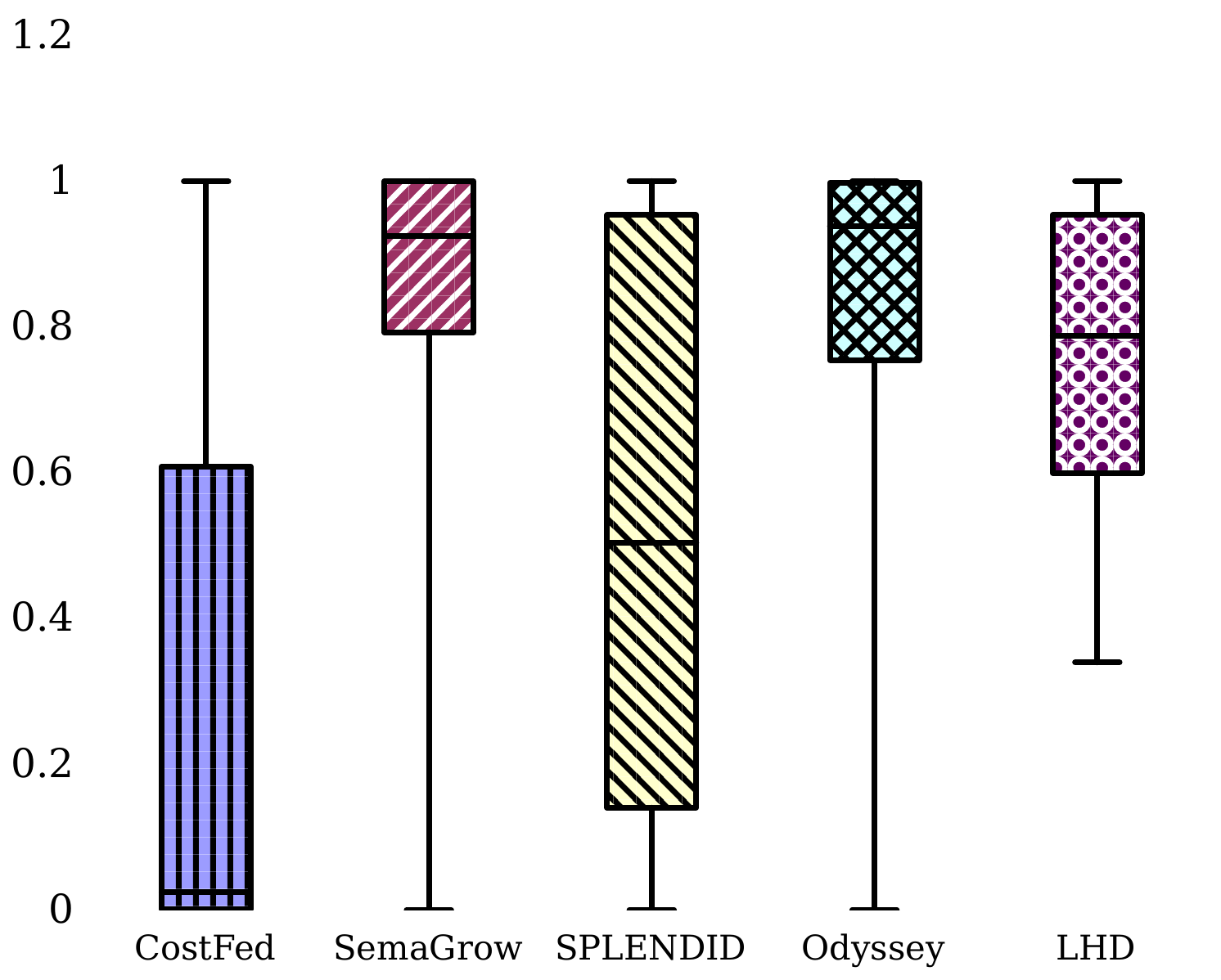}
  \caption{Overall Similarity Error of query plans}
  \label{fig:bgps}
\end{subfigure}%
\begin{subfigure}{.48\textwidth}
  \includegraphics[scale=0.48]{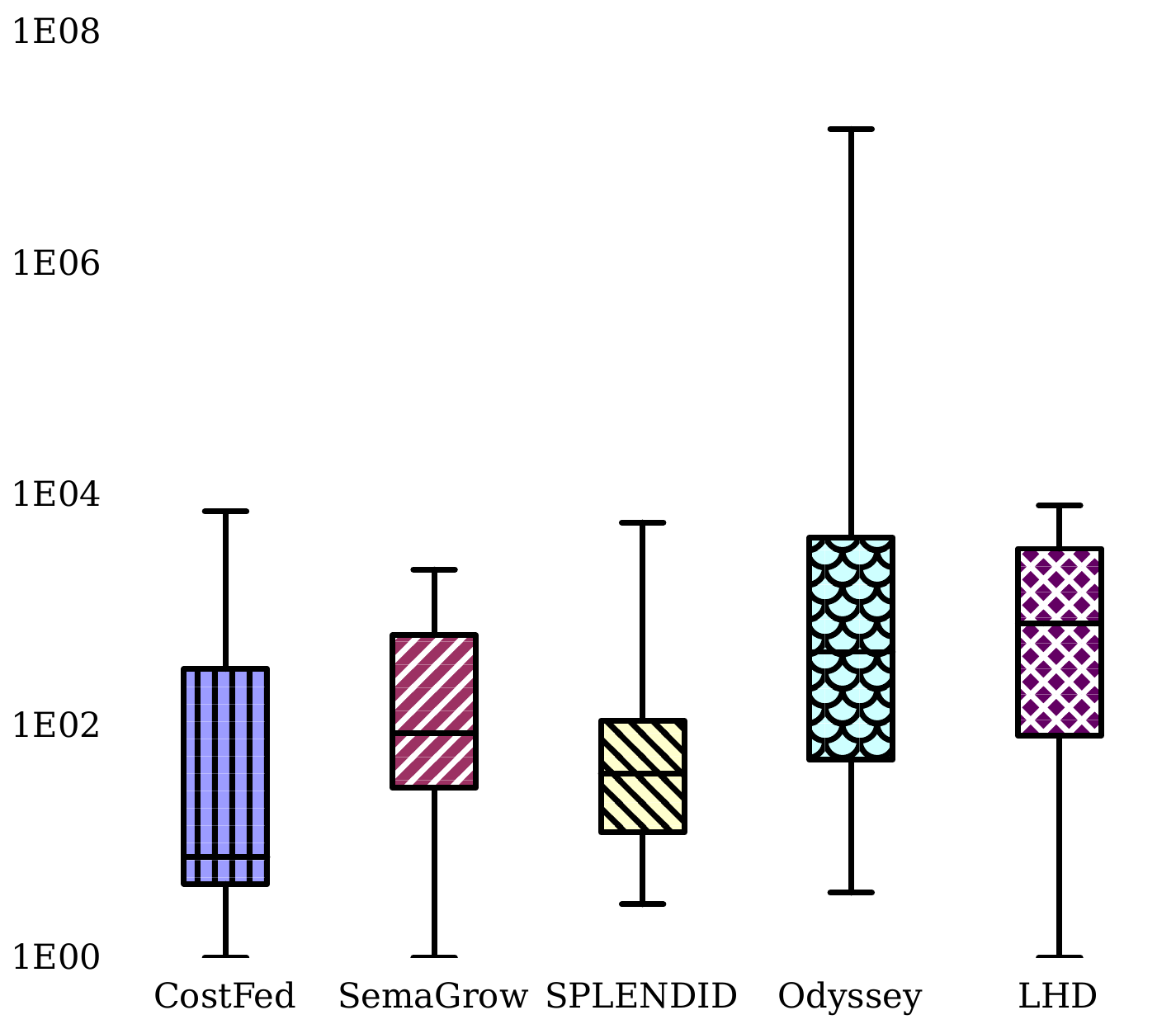}
  \caption{Overall q-error of query plans}
  \label{fig:nlsq}
\end{subfigure}
\begin{subfigure}{.48\textwidth}
  \includegraphics[scale=0.48]{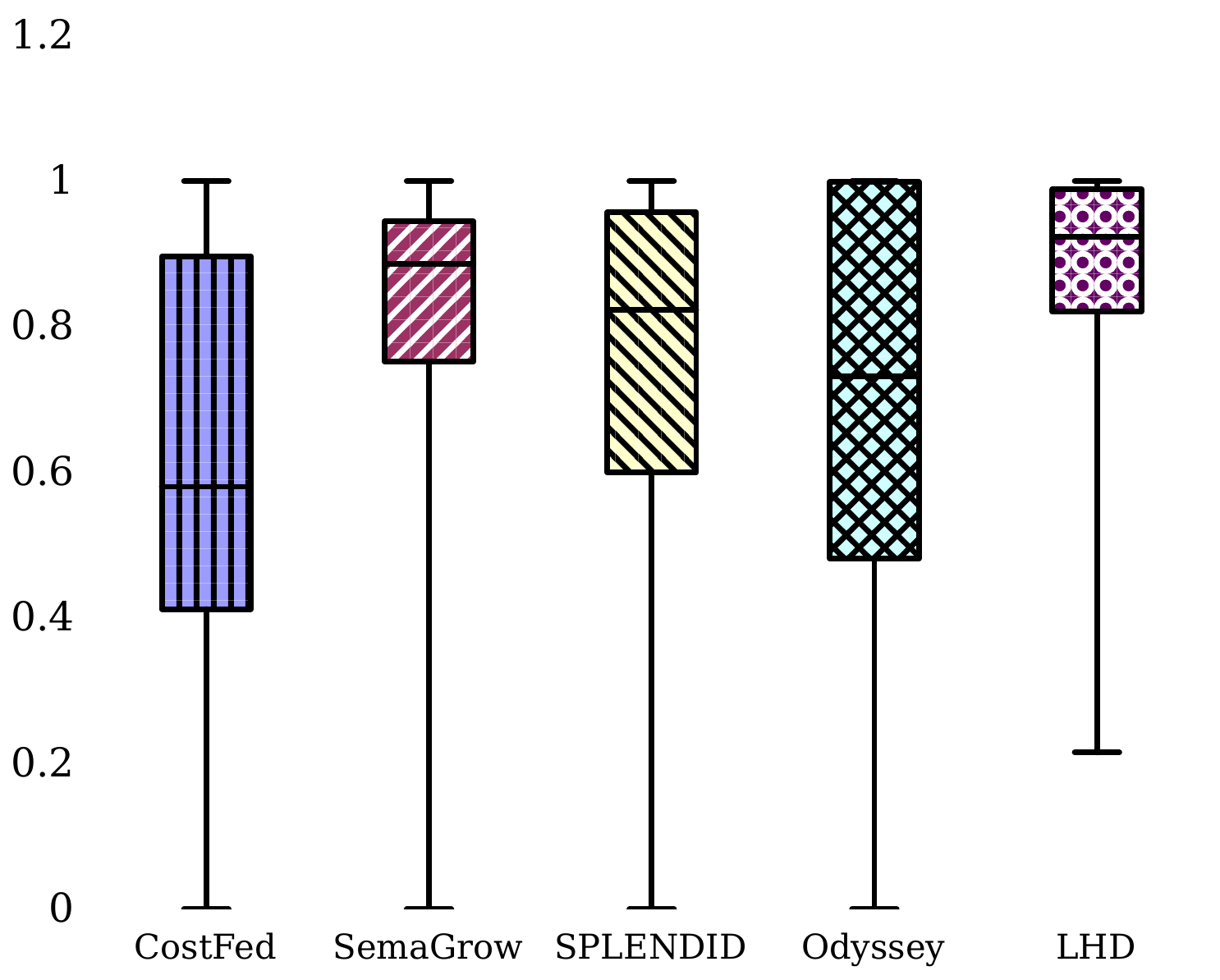}
  \caption{Join Similarity Error of query plans}
  \label{fig:rt}
\end{subfigure}%
\begin{subfigure}{.48\textwidth}
  \includegraphics[scale=0.48]{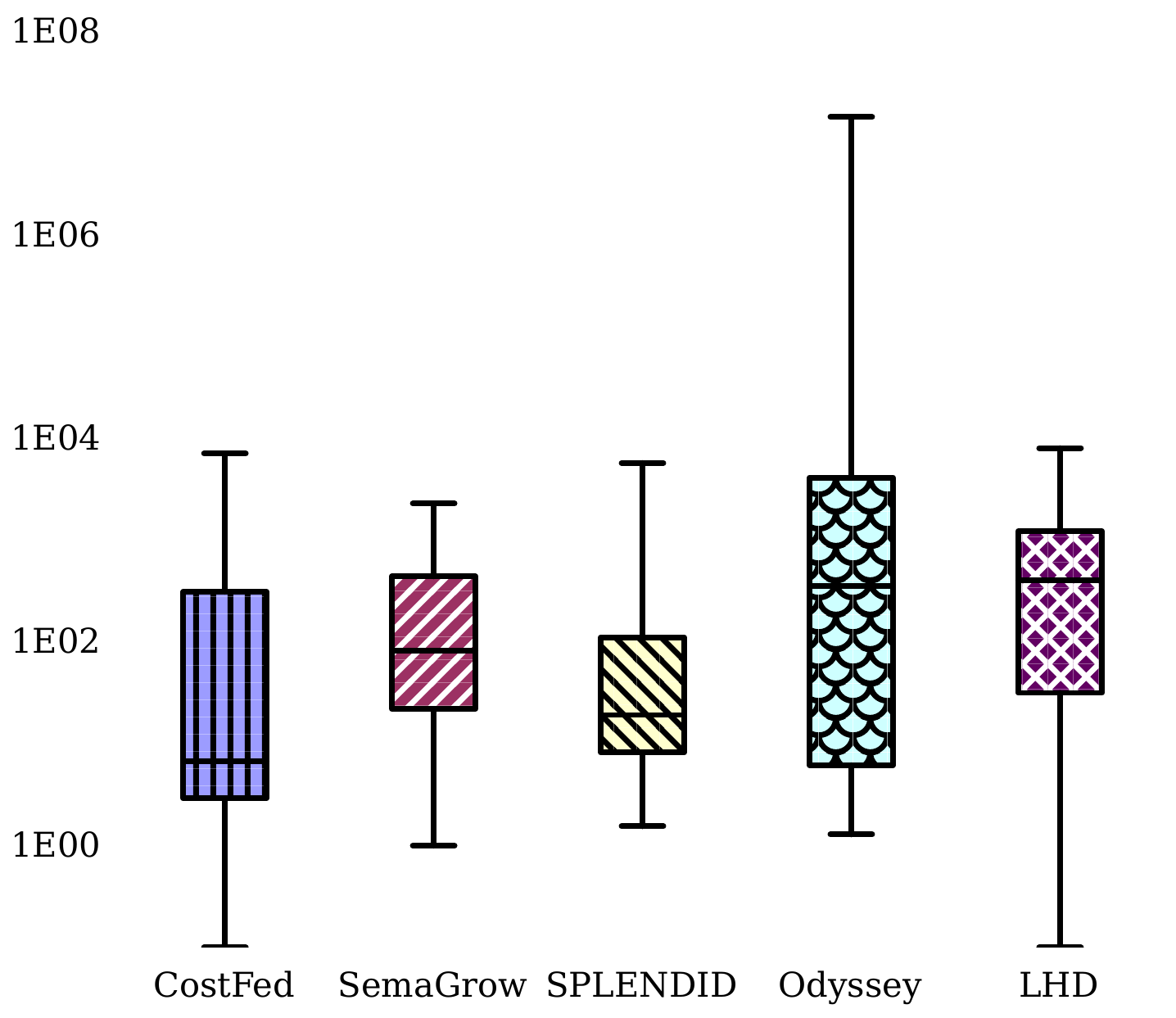}
  \caption{Join q-error of query plan}
  \label{fig:odiversity}
\end{subfigure}
\begin{subfigure}{.48\textwidth}
  \includegraphics[scale=0.48]{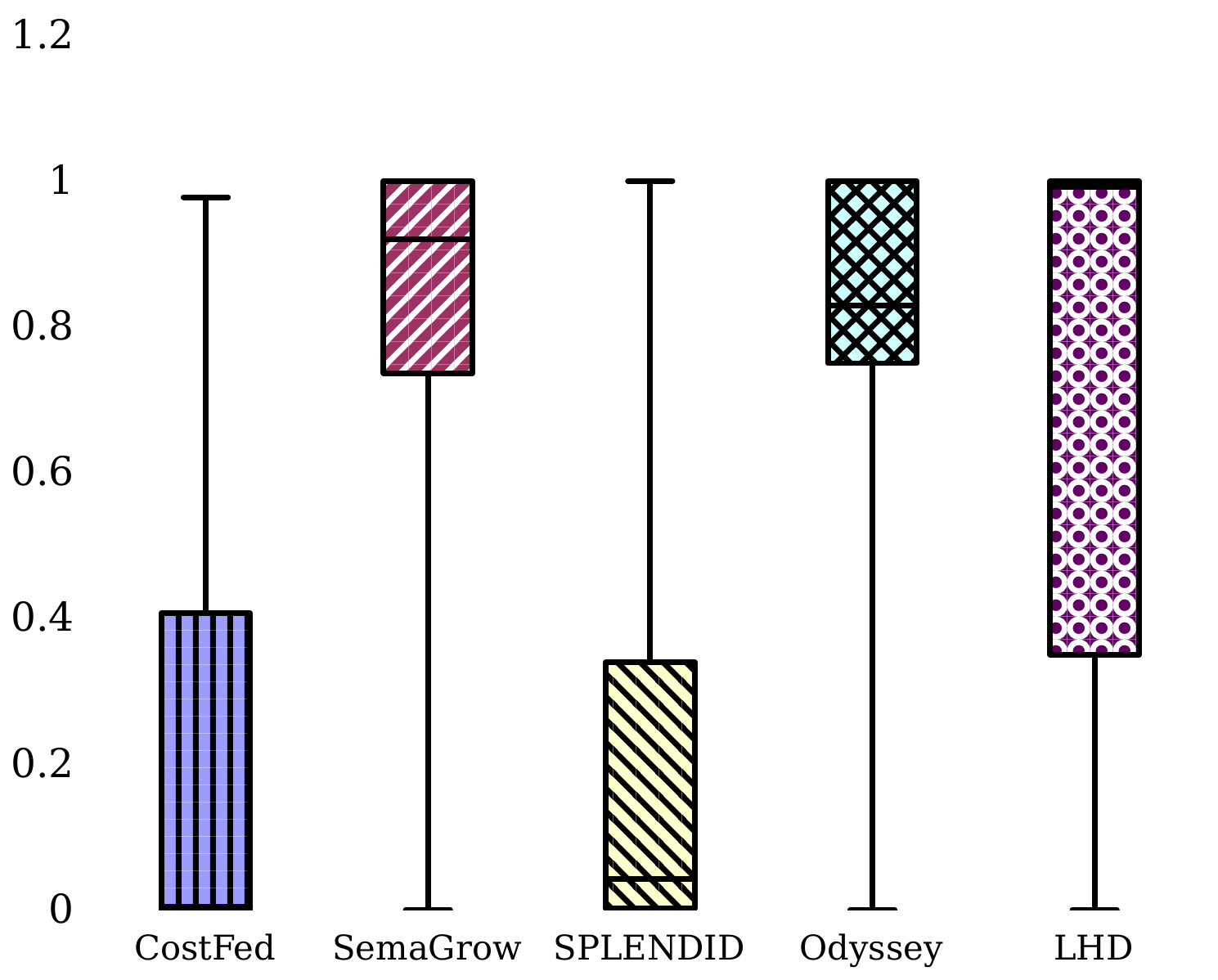}
  \caption{Triple pattern Similarity Error of query}
  \label{fig:rt1}
\end{subfigure}%
\begin{subfigure}{.48\textwidth}
  \includegraphics[scale=0.48]{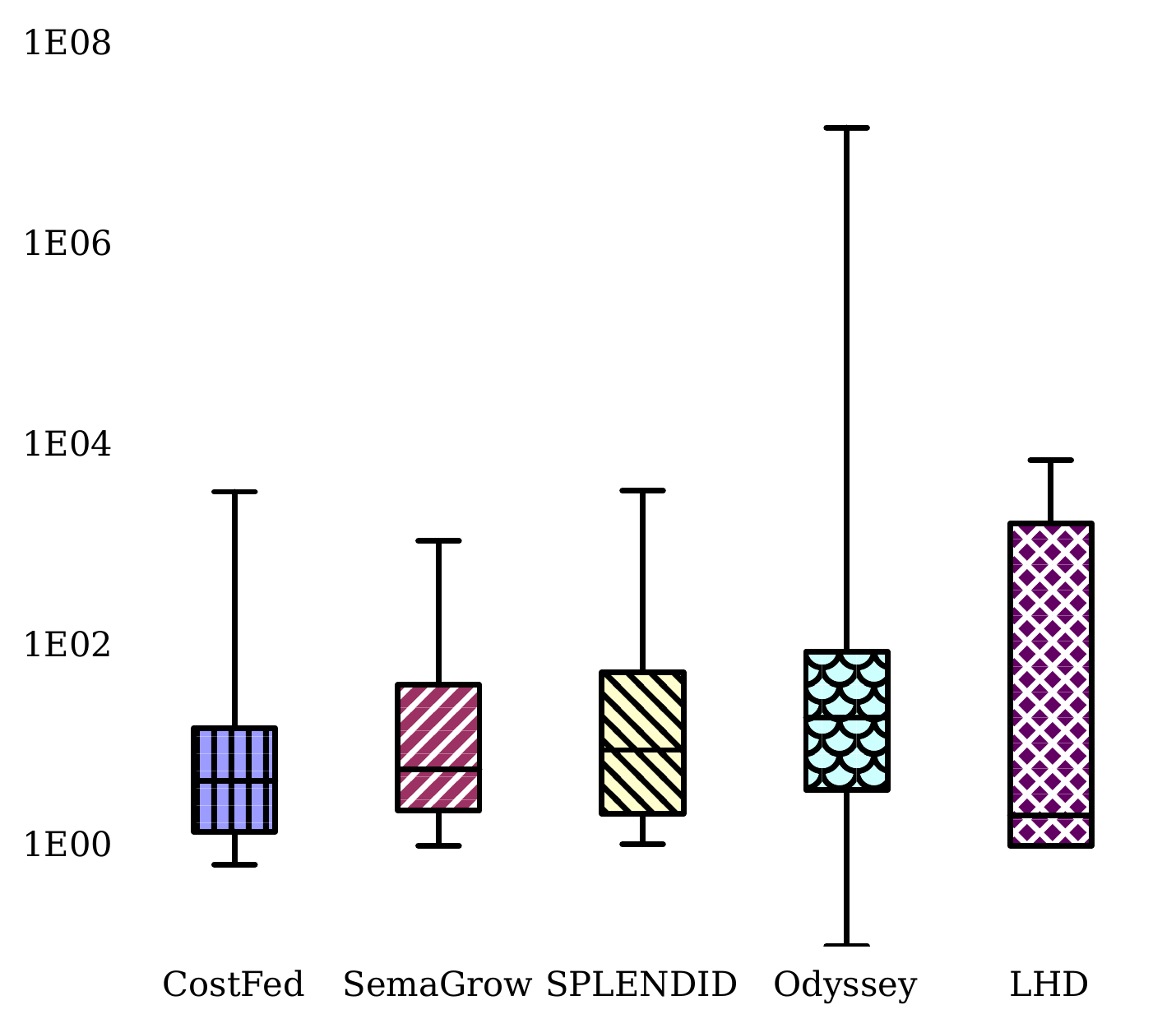}
  \caption{Triple pattern q-error of query}
  \label{fig:odiversity1}
\end{subfigure}
\caption{Similarity and q-error of Query plan}
\label{fig-query-analysis}
\end{figure*}

To further investigate the correlation between metrics and runtimes, we measured Spearman's correlation coefficient between query runtimes and corresponding errors of each of the first six metrics. The results are shown in Table \ref{tab:correlation} which shows that the proposed metrics on average have positive correlations with query runtimes, i.e., the smaller the error, the smaller the query runtimes. The similarity error of overall query plan ($E_P$) has the highest impact (i.e. 0.35) on query runtimes, followed by the similarity error of the triple pattern (i.e., $E_T$ with 0.27), q-error of joins (i.e., $Q_J$ with 0.26), similarity error of Join (i.e., $E_J$ with 0.22), q-error of overall plan (i.e., $Q_P$ with 0.17), and q-error of triple patterns (i.e., $Q_T$ with 0.06). 

In order to make a fair comparison between the results, we only take the common queries on which every system passed. We eliminate the LHD \cite{LHD2013} because it failed in 20/32  benchmark queries (which is a very high number and only 12 simple queries passed), and is not adequate for comparison. We apply Spearman's correlation again on common queries. Table \ref{tab:c_correlation} shows that the proposed metric has a positive correlation with query runtime when we deal with only common queries. The similarity error of overall plan ($E_P$) and triple pattern ($E_T$) has the highest impact (i.e., 0.40) on query runtime, followed by similarity error of joins (i.e., $E_J$ with 0.39), q-error of joins (i.e., $Q_J$ with 0.17) and overall query plan (i.e., $Q_P$ with 0.17), and q-error of triple patterns (i.e., $Q_T$ with 0.01). 

Furthermore, we removed outliers influencing results by applying robust regression on both the q-error and proposed similarity error metrics. Robust regression is done by Iterated Re-weighted Least Squares (IRLS) \cite{IRLS1977}. We used Huber weights \cite{Huber1992} as weighting function in IRLS. This approach further fine-tuned the results and made the correlation for our proposed similarity error and runtime stronger. Table \ref{tab:cr_correlation} shows that all metrics have a positive correlation. However, in our proposed metric this difference is definite. The similarity error of overall query plan ($E_P$) has the highest impact (i.e., 0.56) on query runtimes, followed by the similarity error of the triple pattern (i.e. $E_T$ with 0.49), similarity error of joins ($E_J$ with 0.45), q-error of joins (i.e. $Q_J$ with 0.22), q-error of overall plan (i.e., $Q_P$ with 0.18) and triple pattern (i.e., $Q_P$ with 0.18). Table \ref{tab:cr_correlation} also shows that the q-error for Odyssey is negatively correlated with runtime. We can also observe high q-error values from Figure \ref{fig-query-analysis}. 

Another important factor worth mentioning is that the robust regression does not abide by the normality assumptions. Comparing the p-values (at 5\% confidence level) of the simple linear regression and robust regression suggests that the data is sufficiently  normally distributed for simple linear regression. 

Overall, the results show that the proposed similarity errors correlate better with query runtimes than the q-error. Moreover, the correct estimation of the overall plan is clearly the most crucial fragment of the plan generation. Thus, it is important for federation engines to pay particular attention to the cardinality estimation of the overall query plan. However, given that this estimation commonly depends on triple patterns and join estimations, better means for approximating triple patterns and join cardinalities should lead to better plans. 
The weak to moderate correlation of the similarity errors with query runtimes suggests that the query runtime is a complex measure affected by multi-dimensional metrics, such as metrics given in table \ref{tab:metric-stats} and the SPARQL features, such as number of triple patterns, their selectivities, use of projection variables, number of joins and their types \cite{triplestoreBench2019}. Therefore, it is rather hard to pinpoint a single metric or a SPARQL feature which has a high correlation with the runtime \cite{triplestoreBench2019, LargeRDFBench2018}. The proposed similarity error metric is related to the query planning component of the federation engines and is useful for evaluating the quality of the query plans generated by these engines. 

\subsubsection{Outlier Analysis}
In the robust regression model, the outliers are adjusted with new values according to Huber loss function. In similarity error, the list of queries which are re-weighted after applying robust regression are: C2, C1, S14, CH7 in CostFed; S2 in SemaGrow; S8 and S2 in SPLENDID; and S8 in Odyssey. In q-error, the list of queries which are re-weighted after applying robust regression are: C6, C2, C4, CH7, S3 in CostFed; CH3, CH4, S13, C2 in SemaGrow; CH6, C2, C7, S5 in SPLENDID; and S11, C2, C1, S4 in Odyssey.

In these queries, the residual values are either significantly higher or lower than the regression line. For example, in CostFed the average of the similarity errors across all queries is 0.272 and the range of the residual values for unmodified queries is between -0.17 and 0.17, while C2 similarity error is 0.99 with residual value 0.73, CH7 similarity error is 0.99 with residual value 0.19, and C1 similarity error is 0.62  with residual value 0.32. Hence, by re-weighting these queries, the similarity error and q-error values are re-adjusted close to the regression line, to get a more clear and concise picture of the regression experiments. As it can be observed from the Simple Linear Regression figure, the outliers are influencing the results.  For example, in CostFed, the R value in simple regression is 0.59 and after re-weighting in robust regression the value increased to 0.66. Furthermore, we observe that, in similarity error the R values are increased in robust regression as compared to simple linear regression. While on the other hand, for q-error the R values are decreased in robust regression, further suggesting that similarity error is the better predictor of the runtime as compared to q-error. 

%
Finally, the overall q-error is more affected by robust regression as compared to similarity error. This is because a q-error  takes the maximum of all the errors in the cardinality estimation of the joins and triple patterns. Consequently, some queries produce very high q-error values due to a single less efficient cardinality estimation for a join or a triple pattern.   

\subsubsection{Combined Regression-based Comparison Analysis:}
Recall that our null hypothesis was that there is no correlation between query runtime and error measurement. Based on the results shown in Figures \ref{fig:mvsr1} and \ref{fig:mvsr2}, we can make the following observations:
\begin{itemize}
    \item We can reject the null hypothesis in 62.5\% (i.e., 5 out 8) of the experiments for the similarity error while the same can only be done in 12.5\% (1 out of 8) experimental settings for the q-error. 
    \item The similarity error is significantly correlated with the runtimes of CostFed (simple and robust regression), SemaGrow (simple and robust regression) and Splendid (robust regression). On the other hand, the q-error is solely significantly correlated with the runtime of SemaGrow (robust regression). In the one case where the p-values achieved by both measures allow to reject the null hypothesis (i.e., for SemaGrow using the robust regression analysis), the R-value of the similarity error is higher than that of the q-error (.56 vs. .53). 
    \item For Odyssey, both the similarity error and q-error were not able to produce significant results in our experiments. This suggests that the two errors do not capture the phenomena that influence the performance of Odyssey. A deeper look into Odyssey's runtime performance suggests that it performs worst w.r.t. its source selection time (see Table \ref{tab:metric2}), a factor which is not captured by the errors considered herein.

\end{itemize}
 Our observations suggest that the similarity error is more likely to be significantly correlated with the runtime of a federated query engine than the q-error. However, for some systems (like Odyssey in our case) it may not produce significant results. 
 Interestingly, the correlation between similarity error and runtimes is significant and highest for best-performing (in terms of average query runtime, see Figure~\ref{fig:b1-b2_srcruntime}) federated query engine CostFed. We hypothesize that this result might indicate that the similarity error is most useful for systems which are already optimized to generate good plans. However, this hypothesis needs to be confirmed through further experiments. 
Still, the usefulness of the similarity error seems especially evident when one compares the behaviour of the similarity and the q-error when faced with single cardinality estimation errors. For example, suppose we have 3 joins in a query with estimated cardinalities 10, 10, 100 and with real cardinalities 10, 10 and 1 respectively. The q-error of the plan would be 100 even though only a single join estimation was not optimal. As shown by the equation in Section \ref{sec:qerror}, the q-error is sensitive to single estimation error if they are of high magnitude. This is not the case with similarity errors, which would return 0.86.
   

\subsection{q-error and Similarity-Based Errors}
We now present a comparison of the selected cost-based engines based on the 6 metrics given in Figure \ref{fig-query-analysis}. 
Overall, the similarity errors of query plans given in Figure \ref{fig:bgps} suggests that CostFed produces the smallest errors followed by SPLENDID, LHD, SemaGrow, and Odyssey. CostFed produces smaller errors than SPLENDID in 10/17 comparable queries (excluding queries with timeout and runtime errors). SPLENDID produces smaller errors than LHD in 12/14 comparable queries. LHD produces smaller errors than SemaGrow in 6/12 comparable queries. In turn, SemaGrow produces smaller errors than Odyssey in 9/15 comparable queries.

An overall evaluation of the q-error of query plans given in Figure \ref{fig:nlsq} leads to the following result: CostFed produces the smallest errors followed by SPLENDID, SemaGrow, Odyssey, and LHD. In particular, CostFed produces smaller errors than SPLENDID in 9/17 comparable queries (excluding queries with timeout and runtime error). SPLENDID produces smaller errors 
than SemaGrow in 9/17 comparable queries. 
 SemaGrow produces smaller errors than Odyssey in 8/13 comparable queries. Odyssey is superior to LHD in 5/8 cases.  

An overall evaluation of the similarity error in joins leads to a different picture (see  Figure \ref{fig:rt}). While CostFed remains the best system and produces the smallest errors, it is  followed by Odyssey, SPLENDID, SemaGrow, and LHD. In particular, CostFed outperforms Odyssey in 12/17 comparable queries (excluding queries with timeout and runtime error). Odyssey produces less errors than SPLENDID in 7/14 comparable queries. SPLENDID is superior to SemaGrow in 11/17 comparable queries. SemaGrow outperforms LHD in 7/12 comparable queries.

As an overall evaluation of the q-error of joins given in Figure \ref{fig:odiversity}, CostFed produces the smallest errors followed by SPLENDID, SemaGrow, Odyssey, and LHD. CostFed produces less errors 
than SPLENDID in 12/17 comparable queries (excluding queries with timeout and runtime error). SPLENDID produces less errors 
than SemaGrow in 9/17 comparable queries. SemaGrow produces less errors 
than Odyssey in 9/13 comparable queries. Odyssey produces less errors 
than LHD in 4/8 comparable queries.

Overall, the evaluation of the similarity errors of triple patterns given in Figure \ref{fig:rt1} reveals that CostFed produces the smallest errors 
followed by SPLENDID, Odyssey, SemaGrow, and LHD. CostFed produces smaller errors 
than SPLENDID in 10/17 comparable queries (excluding queries with timeout and runtime error). SPLENDID produces smaller errors 
than Odyssey in 15/17 comparable queries. Odyssey produces smaller errors 
 than SemaGrow in 7/14 comparable queries. SemaGrow outperformed LHD in 6/12 queries. 

An overall evaluation of the \texttt{q-}error of triple patterns given in Figure \ref{fig:odiversity1} leads to a different ranking: CostFed produces the smallest errors followed by LHD, SemaGrow, SPLENDID, and Odyssey. CostFed outperforms LHD in 6/11 comparable queries (excluding queries with timeout and runtime error). LHD produces fewer errors than SemaGrow in 5/10 comparable queries. SemaGrow is better than SPLENDID in 10/17 comparable queries. SPLENDID produces fewer errors 
than Odyssey in 7/14 comparable queries.

In general, the accuracy of the estimation is dependent upon the detail of the statistics stored in the index or data summaries. Furthermore, it is important to pay special attention to the different types of triple patterns (with bound and unbound subject, predicate, objects) and joins types (subject-subject, subject-object, object-object) for the better cardinality estimations. CostFed is more accurate because of the more detailed data summaries, able to handle the different types of triple patterns and joins between triple patterns. The use of the buckets can more accurately estimate the cardinalities of the triple patterns with most common predicates used in the dataset. Furthermore, it handles multi-valued predicates. The Odyssey statistics are more detailed as compared to SPLENDID and SemaGrow (both using VoiD statististics). The distributed characteristic sets (CS) and characteristic pair (CP) statistics generally leads to better cardinality estimations for joins.


\subsection{How Much Does An Efficient Cardinality Estimation Really Matter?}

\definecolor{ne}{HTML}{FF6347}  
\definecolor{nd}{HTML}{32CD32}  
\definecolor{nc}{HTML}{87CEFA}   
\definecolor{nb}{HTML}{D3D3D3}   

\newcommand{\redbox}[1]{\cellcolor{ne}{$#1$}}
\newcommand{\greenbox}[1]{\cellcolor{nd}{$#1$}}
\newcommand{\bluebox}[1]{\cellcolor{nc}{$#1$}}
\newcommand{\greybox}[1]{\cellcolor{nb}{$#1$}}

\begin{table*}[!htb]
 \setlength\tabcolsep{20pt}

\centering
\footnotesize

\begin{tabular}{llPPPPP}
	\toprule  
	&Query &	
	CostFed	 &	SemaGrow&	SPLENDID &  Odyssey & LHD \\
	\midrule
 	\parbox[t]{2mm} {\multirow{14}{*}{\rotatebox[origin=c]{90}{\textbf{Simple Queries}}}}

	& S1 &  	 \bluebox{\texttt{OnlyP}} & 	\bluebox{\texttt{OnlyP}} & 	\bluebox{\texttt{OnlyP}} & 	\bluebox{\texttt{OnlyP}} & 	\bluebox{\texttt{OnlyP}} \\
    & S2 &  	\bluebox{\texttt{OnlyP}} & 	\bluebox{\texttt{OnlyP}} & 	\bluebox{\texttt{OnlyP}} & 	\bluebox{\texttt{OnlyP}} & 	\greenbox{\texttt{OptP}} \\
    & S3 &  	\bluebox{\texttt{OnlyP}} & 	\bluebox{\texttt{OnlyP}} & 	\bluebox{\texttt{OnlyP}} & 	\bluebox{\texttt{OnlyP}} & 	\greenbox{\texttt{OptP}} \\
    & S4 &  	\bluebox{\texttt{OnlyP}} & 	\greenbox{\texttt{OptP}} & 	\bluebox{\texttt{OnlyP}} & 	\bluebox{\texttt{OnlyP}} & 	\greenbox{\texttt{OptP}} \\
    & S5 &  	\bluebox{\texttt{OnlyP}} & 	\greenbox{\texttt{OptP}} & 	\bluebox{\texttt{OnlyP}} & 	\bluebox{\texttt{OnlyP}} & 	\greenbox{\texttt{OptP}} \\
    & S6 &  	\greenbox{\texttt{OptP}} & 	\greenbox{\texttt{OptP}} & 	\greenbox{\texttt{OptP}} & 	\redbox{\texttt{subOpt}} & 	\greenbox{\texttt{OptP}} \\
    & S7 &  	\greenbox{\texttt{OptP}} & 	\greenbox{\texttt{OptP}} & 	\greenbox{\texttt{OptP}} & 	\redbox{\texttt{subOpt}} & 	\greenbox{\texttt{OptP}} \\
    & S8 &  	\bluebox{\texttt{OnlyP}} & 	\bluebox{\texttt{OnlyP}} & 	\bluebox{\texttt{OnlyP}} & 	\bluebox{\texttt{OnlyP}} & 	\bluebox{\texttt{OnlyP}} \\
    & S9 &  	\bluebox{\texttt{OnlyP}} & 	\bluebox{\texttt{OnlyP}} & 	\bluebox{\texttt{OnlyP}} & 	\bluebox{\texttt{OnlyP}} & 	\bluebox{\texttt{OnlyP}} \\ 
    & S10 &  	\bluebox{\texttt{OnlyP}} & 	\redbox{\texttt{subOpt}} & 	\bluebox{\texttt{OnlyP}} & 	\greenbox{\texttt{OptP}} & 	\redbox{\texttt{subOpt}} \\
    & S11 &  	\bluebox{\texttt{OnlyP}} & 	\bluebox{\texttt{OnlyP}} & 	\bluebox{\texttt{OnlyP}} & 	\bluebox{\texttt{OnlyP}} & 	\redbox{\texttt{subOpt}} \\
    & S12 &  	\greenbox{\texttt{OptP}} & 	\greenbox{\texttt{OptP}} & 	\greenbox{\texttt{OptP}} & 	\greenbox{\texttt{OptP}} & 	\redbox{\texttt{subOpt}} \\
    & S13 &  	\bluebox{\texttt{OnlyP}} & 	\greenbox{\texttt{OptP}} & 	\greenbox{\texttt{OptP}} & 	\greenbox{\texttt{OptP}} & 	\redbox{\texttt{subOpt}} \\
    & S14 &  	\bluebox{\texttt{OnlyP}} & 	\greenbox{\texttt{OptP}} & 	\greenbox{\texttt{OptP}} & 	\bluebox{\texttt{OnlyP}} & 	\greenbox{\texttt{OptP}} \\
  \midrule
   \parbox[t]{2mm} {\multirow{10}{*}{\rotatebox[origin=c]{90}{\textbf{Complex Queries}}}}  
    & C1 &  	\bluebox{\texttt{OnlyP}} & 	\greenbox{\texttt{OptP}} & 	\greenbox{\texttt{OptP}} & 	\bluebox{\texttt{OnlyP}} & 	\greybox{\texttt{Failed}} \\
    & C2 &  	\redbox{\texttt{subOpt}} & 	\redbox{\texttt{subOpt}} & 	\redbox{\texttt{subOpt}} & 	\greenbox{\texttt{OptP}} & 	\greybox{\texttt{Failed}} \\
    & C3 &  	\greenbox{\texttt{OptP}} & 	\redbox{\texttt{subOpt}} & 	\greenbox{\texttt{OptP}} & 	\greenbox{\texttt{OptP}} & 	\greybox{\texttt{Failed}} \\
    & C4 &  	\greenbox{\texttt{OptP}} & 	\redbox{\texttt{subOpt}} & 	\redbox{\texttt{subOpt}} & 	\redbox{\texttt{subOpt}} & 	\greybox{\texttt{Failed}} \\
    & C5 &  	\redbox{\texttt{subOpt}} & 	\greenbox{\texttt{OptP}} & 	\redbox{\texttt{subOpt}} & 	\redbox{\texttt{subOpt}} & 	\greybox{\texttt{Failed}} \\
    & C6 &  	\bluebox{\texttt{OnlyP}} & 	\greenbox{\texttt{OptP}} & 	\redbox{\texttt{subOpt}} & 	\bluebox{\texttt{OnlyP}} & 	\greybox{\texttt{Failed}} \\
    & C7 &  	\bluebox{\texttt{OnlyP}} & 	\greenbox{\texttt{OptP}} & 	\greenbox{\texttt{OptP}} & 	\bluebox{\texttt{OnlyP}} & 	\greybox{\texttt{Failed}} \\
    & C8 &  	\bluebox{\texttt{OnlyP}} & 	\redbox{\texttt{subOpt}} & 	\redbox{\texttt{subOpt}} & 	\bluebox{\texttt{OnlyP}} & 	\greybox{\texttt{Failed}} \\
    & C9 &  	\bluebox{\texttt{OnlyP}} & 	\greenbox{\texttt{OptP}} & 	\redbox{\texttt{subOpt}} & 	\bluebox{\texttt{OnlyP}} & 	\greybox{\texttt{Failed}} \\
    & C10 &  	\greenbox{\texttt{OptP}} & 	\greenbox{\texttt{OptP}} & 	\redbox{\texttt{subOpt}} & 	\redbox{\texttt{subOpt}} & 	\greybox{\texttt{Failed}} \\
    \midrule    
    \parbox[t]{2mm} {\multirow{8}{*}{\rotatebox[origin=c]{90}{\textbf{Complex + High Data}}}}  
    {\multirow{8}{*}{\rotatebox[origin=c]{90}{\textbf{Sources Queries}}}}  
    & CH1 &  	\greenbox{\texttt{OptP}} & 	\redbox{\texttt{subOpt}} & 	\redbox{\texttt{subOpt}} & 	\greenbox{\texttt{OptP}} & 	\greybox{\texttt{Failed}} \\
    & CH2 &  	\redbox{\texttt{subOpt}} & 	\redbox{\texttt{subOpt}} & 	\redbox{\texttt{subOpt}} & 	\redbox{\texttt{subOpt}} & 	\greybox{\texttt{Failed}} \\
    & CH3 &  	\greenbox{\texttt{OptP}} & 	\redbox{\texttt{subOpt}} & 	\redbox{\texttt{subOpt}} & 	\redbox{\texttt{subOpt}} & 	\greybox{\texttt{Failed}} \\
    & CH4 &  	\redbox{\texttt{subOpt}} & 	\redbox{\texttt{subOpt}} & 	\redbox{\texttt{subOpt}} & 	\greybox{\texttt{Failed}} & 	\greybox{\texttt{Failed}} \\
    & CH5 &  	\greybox{\texttt{Failed}} & 	\redbox{\texttt{subOpt}} & 	\redbox{\texttt{subOpt}} & 	\redbox{\texttt{subOpt}} & 	\greybox{\texttt{Failed}} \\
    & CH6 &  	\greybox{\texttt{Failed}} & 	\greybox{\texttt{Failed}} & 	\greenbox{\texttt{OptP}} & 	\redbox{\texttt{subOpt}} & 	\greybox{\texttt{Failed}} \\
    & CH7 &  	\redbox{\texttt{subOpt}} & 	\redbox{\texttt{subOpt}} & \redbox{\texttt{subOpt}}	 & 	\redbox{\texttt{subOpt}} & 	\greybox{\texttt{Failed}} \\
    & CH8 &  	\redbox{\texttt{subOpt}} & 	\redbox{\texttt{subOpt}} & 	\redbox{\texttt{subOpt}} & 	\redbox{\texttt{subOpt}} & 	\greybox{\texttt{Failed}} \\
  

 \bottomrule
\end{tabular}
\caption{Query Plans generated by query engines for all queries (Simple, Complex, Complex + High Dimensional Queries).  \textbf{Failed:}(\textcolor{nb}{$\CIRCLE$}) Engine Failed to produce Query Plan,
\textbf{OptP:}(\textcolor{ne}{$\CIRCLE$}) Optimal Query Plan generated by engine,
\textbf{subOpt:}(\textcolor{nd}{$\CIRCLE$}) subOptimal Plan generated by engine,
\textbf{OnlyP:}(\textcolor{nc}{$\CIRCLE$}) Only Plan possible. 
}
\label{tab:plan}
\end{table*}

We observed that it is possible for a federation engine to produce quite a high cardinality estimation error (e.g., 0.99 is the overall similarity error for the S11 query in SemaGrow), yet it produces the optimal query plan. This leads to the question, how much does the efficiency of cardinality estimators of federation engines matter to generate  optimal query plans? To this end, we analyzed query plans generated by each of the selected engines for the benchmark queries. In our analysis, there are three possible cases in each plan:
\begin{itemize}
  
\item \textbf{Optimal plan}: In the \emph{optimal} plan, the best possible join order is selected \emph{based on the given source selection performed by the underlying federation engine}, i.e., the least cardinality joins are always executed first. 

\item \textbf{Sub-optimal plan}: In the \emph{sub-optimal} plan, the engine fails to select the best join \emph{based on the given source selection performed by the underlying federation engine}, i.e., the least cardinality joins are not always executed first. Please note that this also means that the high error in the join cardinality estimation leads to the sub-optimal join order. 

\item \textbf{Only-plan}: In \emph{only-plan}, there is only one possible join order \emph{according to the given source selection performed by the underlying federation engine}. This is possible if only 1 join (excluding a left-join due to the OPTIONAL clause in the query) needs to be executed locally by the federation engine. This situation occurs if there is only a single join in the query or the federation engine creates exclusive groups of joins that are executed remotely by the underlying SPARQL endpoints. 
\end{itemize}

Table \ref{tab:plan} shows the query plan generated by the query planners of the selected engines according to the aforementioned three cases possible for each plan. Since LHD failed to generate any query plan for the majority of the LargeRDFBench queries, we omit it from further discussion. In our evaluation, CostFed produced the smallest sub-optimal plans (i.e, 6) followed by Odyssey (i.e., 11), SemaGrow (i.e., 12), and SPLENDID (i.e, 14). The reason for CostFed's small number of sub-optimal plans is due to the fact that it has the fewest cardinality errors in the estimation, as discussed in the previous section. In addition, it generates the highest number of possible only-plans (which can be regarded as optimal plans for the given source selection information). This is because CostFed's source selection is more efficient in terms of the total triple pattern-wise sources selected without losing recall (see Table \ref{tab:metric2}). 

In Table \ref{tab:plan}, we can see that only a few sub-optimal query plans were generated for simple queries. This is due to the fact that simple category queries of the LargeRDFBench contain very few joins (avg. 2.6 \cite{LargeRDFBench2018}) to be executed by the federation engines. Thus, it is relatively easy to find the best join execution order. However, for complex and complex-plus-high-data sources queries, more sub-optimal plans were generated. This is because these queries contain more joins (around 4 joins on avg. \cite{LargeRDFBench2018}), hence a more accurate join cardinality estimation is required to generate the optimal join ordering plan. In conclusion, efficient cardinality estimation is more important for complex queries with more possible join ordering.

\subsection{Number of Transferred Triples}
\begin{table*}[!htb]
 \setlength\tabcolsep{1.75pt}
\begin{tabular}{l cc cc cc cc cc}

\toprule
Queries	&\multicolumn{2}{c}{CostFed}	&\multicolumn{2}{c}{SemaGrow}	&\multicolumn{2}{c}{Odyssey} &\multicolumn{2}{c}{LHD}	&\multicolumn{2}{c}{SPLENDID}	 \\
\cmidrule(lr){2-3} \cmidrule(lr){4-5} \cmidrule(lr){6-7} \cmidrule(lr){8-9} \cmidrule(lr){10-11}
	&\multicolumn{1}{c}{sent} 	&\multicolumn{1}{c}{received}	&\multicolumn{1}{c}{sent} 	&\multicolumn{1}{c}{received}	&\multicolumn{1}{c}{sent} 	&\multicolumn{1}{c}{received}	&\multicolumn{1}{c}{sent} 	&\multicolumn{1}{c}{received}	&\multicolumn{1}{c}{sent} 	&\multicolumn{1}{c}{received}		\\ \midrule

S1	&\greenbox{\texttt{31}}	&\greybox{\texttt{100}}	
    &\redbox{\texttt{49}}	&\redbox{\texttt{111}}	    
    & \greybox{\texttt{47}}&\greybox{\texttt{100}}	
    &\greybox{\texttt{34}}	&\greenbox{\texttt{91}}	
    &\redbox{\texttt{49}}	&\redbox{\texttt{111}}	\\
    
S2	&\greybox{\texttt{12}}&\greybox{\texttt{11}}	    
    &\redbox{\texttt{21}}&\greybox{\texttt{12}}	    
    &\greybox{\texttt{11}}&\greybox{\texttt{11}}	   
    &\greenbox{\texttt{10}}&\greenbox{\texttt{3}}	    
    &\redbox{\texttt{21}}&\redbox{\texttt{12}}	\\
    
S3	&\greenbox{\texttt{20}}&\greenbox{\texttt{2}}	    
    &\greybox{\texttt{22}}	&\redbox{\texttt{24}}	    
    &\greenbox{\texttt{20}}	&\greenbox{\texttt{2}}	   
    &\greybox{\texttt{Failed}}	&\greybox{\texttt{Failed}}	    
    &\redbox{\texttt{25}}	&\greybox{\texttt{4}}	\\
    
S4	&\greenbox{\texttt{12}}	&\greenbox{\texttt{1}}	    
    &\redbox{\texttt{46}}&\greybox{\texttt{17}}	    
    &\greenbox{\texttt{12}}	&\greenbox{\texttt{1}}	    
    &\greybox{\texttt{20}}	&\redbox{\texttt{18}}	    
    &\greybox{\texttt{15}}&\greybox{\texttt{3}}	\\
    
S5	&\greenbox{\texttt{16}}	&\greybox{\texttt{17}}	    
    &\redbox{\texttt{21}}&\greenbox{\texttt{12}}	    
    &\greenbox{\texttt{16}}&\greybox{\texttt{17}}	    
    &\greybox{\texttt{18}}&\redbox{\texttt{13444}}	    
    &\greybox{\texttt{18}}	&\greybox{\texttt{20}}	\\
    
S6	&\greenbox{\texttt{34}}&\greenbox{\texttt{1616}}	    
    &\greybox{\texttt{TO}}&\greybox{\texttt{TO}}	    
    &\redbox{\texttt{6500}}&\redbox{\texttt{3254}}	    
    &\greybox{\texttt{283}}&\greybox{\texttt{1766}}	    
    &\greybox{\texttt{36}}&\greybox{\texttt{1618}}	\\
    
S7	&\greybox{\texttt{27}}	&\redbox{\texttt{642}}	    
    &\redbox{\texttt{45}}&\greybox{\texttt{635}}	
    &\greybox{\texttt{TO}}&\greybox{\texttt{TO}}	    
    &\greenbox{\texttt{20}}&\greenbox{\texttt{143}}	    
    &\greybox{\texttt{29}}&\greybox{\texttt{371}}	\\
    
S8	&\redbox{\texttt{10}}&\greenbox{\texttt{1159}}	    
    &\redbox{\texttt{10}}&\greenbox{\texttt{1159}}	    
    &\redbox{\texttt{10}}&\greenbox{\texttt{1159}}	    
    &\greenbox{\texttt{4}}&\greenbox{\texttt{1159}}	    
    &\redbox{\texttt{10}}&\greenbox{\texttt{1159}}	\\
    
S9	&\greenbox{\texttt{25}}&\redbox{\texttt{351}}	
    &\redbox{\texttt{59}}&\redbox{\texttt{382}}	
    &\greybox{\texttt{25}}&\greenbox{\texttt{351}}	
    &\greybox{\texttt{34}}	&\greybox{\texttt{342}}	
    &\redbox{\texttt{59}}	&\redbox{\texttt{382}}	\\
    
S10	&\greybox{\texttt{20}}&\greybox{\texttt{20054}}	
    &\redbox{\texttt{36}}&\greenbox{\texttt{14578}}	    
    &\greybox{\texttt{TO}}&\greybox{\texttt{TO}}	    
    &\greenbox{\texttt{16}}&\redbox{\texttt{24540}}	    
    &\greybox{\texttt{21}}&\greybox{\texttt{20055}}	\\
    
S11	&\greenbox{\texttt{12}}&\greenbox{\texttt{13}}	    
    &\greybox{\texttt{14}}&\greybox{\texttt{15}}	    
    &\greenbox{\texttt{12}}&\greenbox{\texttt{13}}	    
    &\redbox{\texttt{19}}&\redbox{\texttt{4261}}	
    &\greybox{\texttt{14}}&\greybox{\texttt{15}}	\\
    
S12	&\greenbox{\texttt{2147}}&\greenbox{\texttt{2136}}	    
    &\redbox{\texttt{7772}}&\redbox{\texttt{3428}}	    
    &\greenbox{\texttt{2147}}&\greenbox{\texttt{2136}}	
    &\greybox{\texttt{Failed}}	&\greybox{\texttt{Failed}}	    
    &\greybox{\texttt{7442}}&\redbox{\texttt{3428}}	\\
    
S13	&\greenbox{\texttt{68}}&\greybox{\texttt{228}}	    
    &\greybox{\texttt{1161}}&\greybox{\texttt{10267}}	    
    &\greybox{\texttt{105}}&\greenbox{\texttt{131}}	    
    &\redbox{\texttt{2456}}&\redbox{\texttt{13079}}	    
    &\greybox{\texttt{1161}}&\greybox{\texttt{10267}}	\\
    
S14	&\greybox{\texttt{2877}}&\greybox{\texttt{4033}}	    
    &\redbox{\texttt{3852}}&\redbox{\texttt{4366}}	    
    &\greybox{\texttt{2877}}&\greybox{\texttt{4033}}	    
    &\greenbox{\texttt{97}}&\greenbox{\texttt{2449}}	    
    &\redbox{\texttt{3852}}&\redbox{\texttt{4366}}	\\

\rowcolor{LightCyan2} \textbf{Avg}&
\textbf{\texttt{379}}& \textbf{\texttt{2168}}&
\redbox{\textbf{\texttt{1008}}}& \textbf{\texttt{2693}}&
\textbf{\texttt{981}}& \greenbox{\textbf{\texttt{934}}}&
\greenbox{\textbf{\texttt{250}}}& \redbox{\textbf{\texttt{5108}}}&
\textbf{\texttt{910}}& \textbf{\texttt{2987}}\\

\midrule

C1	&\greenbox{\texttt{4234}}&\greenbox{\texttt{2573}}	    
    &\greybox{\texttt{Failed}}	&\greybox{\texttt{Failed}}	    
    &\greenbox{\texttt{4234}}&\greenbox{\texttt{2573}}	         
    &\greybox{\texttt{Failed}}&\greybox{\texttt{Failed}}	    
    &\redbox{\texttt{5232}}&\redbox{\texttt{4173}}	\\
    
C2	&\redbox{\texttt{1371}}&\redbox{\texttt{2118}}	    
    &\greenbox{\texttt{131}}&\greybox{\texttt{1532}}	        
    &\greybox{\texttt{1352}}&\greenbox{\texttt{1354}}	
    &\greybox{\texttt{Failed}}&\greybox{\texttt{Failed}}	    
    &\greenbox{\texttt{131}}&\greybox{\texttt{1532}}	\\
    
C3	&\greenbox{\texttt{6213}}&\greybox{\texttt{13343}}	    
    &\greybox{\texttt{8670}}&\redbox{\texttt{15464}}	        
    &\greenbox{\texttt{6213}}&\greybox{\texttt{13343}}	           
    &\greybox{\texttt{Failed}}&\greybox{\texttt{Failed}}	    
    &\redbox{\texttt{13854}}&\greenbox{\texttt{9796}}	\\
    
C4	&\redbox{\texttt{26}}&\greenbox{\texttt{550}}	    
    &\greybox{\texttt{TO}}&\greybox{\texttt{TO}}	    
    &\greenbox{\texttt{11}}	&\redbox{\texttt{1093}}	    
    &\greybox{\texttt{Failed}}&\greybox{\texttt{Failed}}	
    &\greybox{\texttt{Failed}}&\greybox{\texttt{Failed}}	\\
    
C5	&\greybox{\texttt{TO}}&\greybox{\texttt{TO}}	    
    &\greybox{\texttt{Failed}}&\greybox{\texttt{Failed}}	    
    &\greenbox{\texttt{2232}}&\greenbox{\texttt{20532}}	      
    &\greybox{\texttt{Failed}}&\greybox{\texttt{Failed}}	    
    &\greybox{\texttt{TO}}&\greybox{\texttt{TO}}	\\
    
C6	&\greenbox{\texttt{12}}&\greenbox{\texttt{11432}}	    
    &\greybox{\texttt{TO}}&\greybox{\texttt{TO}}	    
    &\greenbox{\texttt{12}}&\greenbox{\texttt{11432}}	    
    &\greybox{\texttt{Failed}}&\greybox{\texttt{Failed}}	    
    &\redbox{\texttt{20}}&\redbox{\texttt{125310}}	\\
    
C7	&\greenbox{\texttt{87}}&\greenbox{\texttt{112}}	    
    &\redbox{\texttt{550}}&\greybox{\texttt{335}}	    
    &\greenbox{\texttt{87}}&\greenbox{\texttt{112}}	
    &\greybox{\texttt{Failed}}&\greybox{\texttt{Failed}}	    
    &\redbox{\texttt{550}}&\redbox{\texttt{335}}	\\
    
C8	&\greenbox{\texttt{622}}&\greenbox{\texttt{3519}}	    
    &\greybox{\texttt{1365}}&\redbox{\texttt{4768}}	    
    &\greenbox{\texttt{622}}&\greenbox{\texttt{3519}}	    
    &\greybox{\texttt{Failed}}&\greybox{\texttt{Failed}}	    
    &\redbox{\texttt{1365}}&\redbox{\texttt{4768}}	\\
    
C9	&\redbox{\texttt{7274}}&\redbox{\texttt{21178}}	    
    &\greenbox{\texttt{3358}}&\greenbox{\texttt{10275}}	    
    &\greybox{\texttt{TO}}&\greybox{\texttt{TO}}	    
    &\greybox{\texttt{Failed}}&\greybox{\texttt{Failed}}	    
    &\greybox{\texttt{Failed}}&\greybox{\texttt{Failed}}	\\
    
C10	&\greenbox{\texttt{51}}&\redbox{\texttt{5702}}	    
    &\greybox{\texttt{112}}&\greybox{\texttt{1541}}	    
    &\greybox{\texttt{Failed}}&\greybox{\texttt{Failed}}	    
    &\greybox{\texttt{Failed}}&\greybox{\texttt{Failed}}	    
    &\redbox{\texttt{979}}&\greenbox{\texttt{1312}}	\\

\rowcolor{LightCyan2} \textbf{Avg}& 
\textbf{\texttt{2210}}& \textbf{\texttt{6725}}                                
& \textbf{\texttt{2364}}& \greenbox{\textbf{\texttt{5652}}}                       
& \greenbox{\textbf{\texttt{1845}}}& \textbf{\texttt{6745}}       
& \textbf{\texttt{NA}}& \textbf{\texttt{NA}}                                     
& \redbox{\textbf{\texttt{3164}}}& \redbox{\textbf{\texttt{21032}}}  \\

\midrule

CH1	&\greenbox{\texttt{390}}	&\greenbox{\texttt{8253}}	
    &\redbox{\texttt{1709}}	&\redbox{\texttt{9439}}	
    &\greybox{\texttt{TO}}	&\greybox{\texttt{TO}}	
    &\greybox{\texttt{Failed}}	&\greybox{\texttt{Failed}}	
    &\greybox{\texttt{Failed}}	&\greybox{\texttt{Failed}}	\\
    
CH2	&\greybox{\texttt{TO}}	&\greybox{\texttt{TO}}	
    &\greybox{\texttt{TO}}	&\greybox{\texttt{TO}}	
    & \greybox{\texttt{TO}}	&\greybox{\texttt{TO}}	
    &\greybox{\texttt{Failed}}	&\greybox{\texttt{Failed}}	
    &\greybox{\texttt{Failed}}	&\greybox{\texttt{Failed}}	\\
    
CH3	&\greenbox{\texttt{167}}	&\redbox{\texttt{5053}}	        
    &\redbox{\texttt{4686}}	&\greenbox{\texttt{4011}}	    
    &\greybox{\texttt{TO}}	&\greybox{\texttt{TO}}	
    &\greybox{\texttt{Failed}}	&\greybox{\texttt{Failed}}	
    &\greybox{\texttt{Failed}}	&\greybox{\texttt{Failed}}	\\
    
CH4	&\redbox{\texttt{72}}	&\redbox{\texttt{25}}	    
    &\greenbox{\texttt{39}}	&\greenbox{\texttt{20}}	
    &\greybox{\texttt{Failed}}	&\greybox{\texttt{Failed}}	
    &\greybox{\texttt{Failed}}	&\greybox{\texttt{Failed}}	
    &\greybox{\texttt{Failed}}	&\greybox{\texttt{Failed}}	\\
    
CH5	&\greybox{\texttt{Failed}}	&\greybox{\texttt{Failed}}	
    &\greybox{\texttt{Failed}}	&\greybox{\texttt{Failed}}	
    &\greybox{\texttt{TO}}	&\greybox{\texttt{TO}}	
    &\greybox{\texttt{Failed}}	&\greybox{\texttt{Failed}}	
    &\greybox{\texttt{TO}}	&\greybox{\texttt{TO}}	\\
    
CH6	&\greybox{\texttt{Failed}}	&\greybox{\texttt{Failed}}	
    &\greybox{\texttt{Failed}}	&\greybox{\texttt{Failed}}	
    &\greybox{\texttt{TO}}	&\greybox{\texttt{TO}}	
    &\greybox{\texttt{Failed}}	&\greybox{\texttt{Failed}}	
    &\greenbox{\texttt{1551}}	&\greenbox{\texttt{9401}}	\\
    
CH7	&\greenbox{\texttt{2332}}	&\greenbox{\texttt{85158}}	
    &\greybox{\texttt{Failed}}	&\greybox{\texttt{Failed}}	
    &\greybox{\texttt{TO}}	&\greybox{\texttt{TO}}	
    &\greybox{\texttt{Failed}}	&\greybox{\texttt{Failed}}	
    &\greybox{\texttt{Failed}}	&\greybox{\texttt{Failed}}	\\

CH8	&\greybox{\texttt{TO}}	&\greybox{\texttt{TO}}	
    &\greybox{\texttt{Failed}}	&\greybox{\texttt{Failed}}	
    &\greybox{\texttt{TO}}	&\greybox{\texttt{TO}}	
    &\greybox{\texttt{Failed}}	&\greybox{\texttt{Failed}}	
    &\greybox{\texttt{Failed}}	&\greybox{\texttt{Failed}}	\\
    
\rowcolor{gray} \textbf{Avg}&  
    \greenbox{\textbf{\texttt{740}}} & \redbox{\textbf{\texttt{24622}}} 
    & \redbox{\textbf{\texttt{2145}}} &\greenbox{\textbf{\texttt{4490}}} 
    & \textbf{\texttt{NA}}& \textbf{\texttt{NA}}
    & \textbf{\texttt{NA}}& \textbf{\texttt{NA}}
    & \textbf{\texttt{1551}}& \textbf{\texttt{9401}} \\
\bottomrule
\end{tabular}
\caption{Number of transferred tuples. \textbf{NA:} "Not applicable". \textbf{Failed} means either "Runtime Error" or "Incomplete Results" and \textbf{TO:} "Timeout", which means Query Execution exceeds threshold value. "green color"(\textcolor{nd}{$\CIRCLE$}) means lowest value among all systems, and "red color"(\textcolor{ne}{$\CIRCLE$}) means highest value among all systems.}
\label{tab:transfer}
\end{table*}

Table \ref{tab:transfer} shows the number of tuples sent and received during the query execution for the selected federation engines. The number of sent tuples is related to the number of endpoint requests sent by the federation engine during query processing \cite{Odyssey2017,fedx2011}. The number of received tuples can be regarded as the number of intermediate results produced by the federation engine during query processing \cite{Odyssey2017}. The smaller number of transferred tuples is considered important for fast query processing \cite{Odyssey2017}. In this regard, CostFed ranked first with 31 green boxes (i.e., it had the best results among the selected engines), followed by Odyssey with 24 green boxes, SemaGrow with 12 green boxes, LHD with 10 green boxes, and then SPLENDID with 9 green boxes. 

In most queries, CostFed and Odyssey produced the only possible plans \emph{only-plan}, which means only one (excluding the Left join for OPTIONAL SPARQL operator) was locally executed by the federation engine. Consequently, these engines transfer fewer tuples in comparison to other approaches. The largest difference is observed for S13, where CostFed and Odyssey clearly outperform the other approaches, transferring 500 times fewer tuples. 
The number of received tuples in LHD is significantly high in comparison to other approaches. This is because it does not produce normal tree-like query plans. Rather, LHD focuses on generating independent tasks that can be run in parallel. Therefore, independent tasks retrieve a lot of intermediate results, which need to be joined locally in order to get the final query resultset. 

Another advantage that CostFed and Odyssey have over other approaches is their join-aware approach for triple pattern-wise sources selected (TPWSS). This join-aware nature of these engines saves many tuples from transferring due to less overestimation of sources. CostFed also performs better because it maintains cache for ask requests and saves many queries from sending to different sources. Another important factor worth mentioning here is that the number of transferred tuples does not consider the number of columns (i.e., the number of projection variables in the query) in the result set, but  only counts the number of rows (i.e., the number of results) returned or sent to the endpoints. We also observed that in the case of an \emph{only-plan} or an \emph{optimal} plan, the number of received tuples is less compared to \emph{sub-optimal} plans, clearly indicating that a smaller number of tuples is key to fast query processing. The amalgamated average of all queries could also be misleading because in complex queries, there are more failed/timeout queries for some systems while producing answers in others. Therefore, we calculated the separate average for each category of queries, i.e., simple, complex and complex-and-high-data. From our analysis of the results, we concludes that if an engine produces \emph{optimal} or \emph{only-plan}, the number of intermediate results also decreases.

\begin{table*}
 \caption{Comparison of index construction time (\textbf{Index Gen.  Time}) and  \textbf{Index Size} for selected federation engines }
 \label{tab:indx}
\begin{tabular}{lccccc} \hline
	&       CostFed& SemaGrow&	SPLENDID&	Odyssey&	LHD \\\hline
Index Gen.  Time (min) &	    65&	    110&	    110&	    533&	110\\
Index Size (MBs)&	  10&	    1&	        1&	        5200 &     1\\

\hline
\end{tabular}
\end{table*}

\begin{table*}[!t]
\caption{Comparison of selected federation engines in terms of source selection time \textbf{ST} in msec, total number of SPARQL \texttt{ASK} requests \textbf{\#A}, and total triple pattern-wise sources selected \textbf{\#T}. (\textbf{RE} represents "Runtime Error",\textbf{TO} represents "Time Out" of 20 min, \textbf{T/A} represents "Total/Average"  where Average is for ST, and Total is for \#T and \#A, \textbf{NA} represents "Not Applicable"). "green color"(\textcolor{nd}{$\CIRCLE$}) means lowest value among all systems, and "red color"(\textcolor{ne}{$\CIRCLE$}) means highest value among all systems.}
\centering
\begin{tabular}{b ccc ccc ccc ccc ccc} 
\toprule
&\multicolumn{3}{c}{Odyssey}& \multicolumn{3}{c}{SPLENDID}&  \multicolumn{3}{c}{LHD}& \multicolumn{3}{c}{SemaGrow}&\multicolumn{3}{c}{CostFed}\\
\cmidrule(lr){2-4} \cmidrule(lr){5-7} \cmidrule(lr){8-10} \cmidrule(lr){11-13} \cmidrule(lr){14-16}

Qry&       \#T&    \#A&   ST	  &          \#T&     \#A&     ST  &           \#T&            \#A&       ST &       \#T&   \#A&   ST  &\#T &\#A &ST  \\

\midrule

S1&	    \greybox{\texttt{11}}&   \greenbox{\texttt{0}}&   \greenbox{\texttt{1}}&	 	    
        \greybox{\texttt{11}}&	\redbox{\texttt{26}}&	\redbox{\texttt{293}}&	
        \redbox{\texttt{28}}&	\greenbox{\texttt{0}}&	\greybox{\texttt{261}}&        
        \greybox{\texttt{11}}&   \redbox{\texttt{26}}&	\redbox{\texttt{293}}&    	
        \greenbox{\texttt{4}}&	\greybox{\texttt{18}}&	\greybox{\texttt{6}}\\
S2&	    \greenbox{\texttt{3}}&	\greenbox{\texttt{0}}&	\greybox{\texttt{14}}&	  	    
        \greenbox{\texttt{3}}&	\redbox{\texttt{9}}&	    \redbox{\texttt{33}}&	 	
        \redbox{\texttt{10}}&	\greenbox{\texttt{0}}&	\greybox{\texttt{8}}&	     	
        \greenbox{\texttt{3}}&	\redbox{\texttt{9}}&	    \redbox{\texttt{33}}&	        
        \greenbox{\texttt{3}}&	\redbox{\texttt{9}}&	    \greenbox{\texttt{1}}\\
        
S3&	    \greenbox{\texttt{5}}&	\greenbox{\texttt{0}}&	\redbox{\texttt{44}}&	 	    
        \greybox{\texttt{12}}&	\redbox{\texttt{2}}&	    \greybox{\texttt{17}}&		
        \redbox{\texttt{20}}&	\greenbox{\texttt{0}}&	\greybox{\texttt{34}}&	 	    
        \greybox{\texttt{12}}&	\redbox{\texttt{2}}&	    \greybox{\texttt{17}}&	    	
        \greenbox{\texttt{5}}&	\greenbox{\texttt{0}}&	\greenbox{\texttt{1}}\\
        
S4&	    \greenbox{\texttt{5}}&	\greenbox{\texttt{0}}&	\redbox{\texttt{321}}&	 	
        \greybox{\texttt{19}}&	\redbox{\texttt{2}}&	    \greybox{\texttt{14}}&		
        \redbox{\texttt{20}}&	\greenbox{\texttt{0}}&	\greybox{\texttt{15}}&	     	
        \greybox{\texttt{19}}&	\redbox{\texttt{2}}&	    \greybox{\texttt{14}}&	    	
        \greenbox{\texttt{5}}&	\greenbox{\texttt{0}}&	\greenbox{\texttt{1}}\\

S5&	    \greenbox{\texttt{4}}&	\greenbox{\texttt{0}}&	\redbox{\texttt{223}}&		
        \redbox{\texttt{11}}&	\redbox{\texttt{1}}&	    \greybox{\texttt{11}}&	 	
        \redbox{\texttt{11}}&	\greenbox{\texttt{0}}&	\greybox{\texttt{8}}&	     	
        \redbox{\texttt{11}}&	\redbox{\texttt{1}}&	    \greybox{\texttt{11}}&	    	
        \greenbox{\texttt{4}}&	\greenbox{\texttt{0}}&	\greenbox{\texttt{1}}\\
        
S6&	    \greenbox{\texttt{6}}&	\greenbox{\texttt{0}}&	\redbox{\texttt{88}}&  	    
        \greybox{\texttt{9}}&	\redbox{\texttt{2}}&	    \greybox{\texttt{16}}&		
        \redbox{\texttt{10}}&	\greenbox{\texttt{0}}&	\greybox{\texttt{36}}&	    	
        \greybox{\texttt{9}}&	\redbox{\texttt{2}}&	    \greybox{\texttt{16}}&	    	
        \greybox{\texttt{8}}&	\greenbox{\texttt{0}}&	\greenbox{\texttt{3}}\\
        
S7&	    \greenbox{\texttt{6}}&	\greenbox{\texttt{0}}&	\redbox{\texttt{72}}&  	    
        \redbox{\texttt{13}}&	\redbox{\texttt{2}}&	    \greybox{\texttt{19}}&		
        \redbox{\texttt{13}}&	\greenbox{\texttt{0}}&	\greybox{\texttt{67}}&	    	
        \redbox{\texttt{13}}&	\redbox{\texttt{2}}&	    \greybox{\texttt{19}}&	    	
        \greenbox{\texttt{6}}&	\greenbox{\texttt{0}}&	\greenbox{\texttt{1}}\\
        
S8&	    \greenbox{\texttt{1}}&	\greenbox{\texttt{0}}&	\redbox{\texttt{8}}&	 	    
        \greenbox{\texttt{1}}&	\greenbox{\texttt{0}}&	\greybox{\texttt{2}}&	 	
        \greenbox{\texttt{1}}&	\greenbox{\texttt{0}}&	\greybox{\texttt{5}}&	     	
        \greenbox{\texttt{1}}&	\greenbox{\texttt{0}}&	\greybox{\texttt{2}}&	        
        \greenbox{\texttt{1}}&	\greenbox{\texttt{0}}&	\greenbox{\texttt{1}}\\
        
S9&	    \greenbox{\texttt{4}}&	\greenbox{\texttt{0}}&	\greenbox{\texttt{1}}&	 	    
        \greybox{\texttt{11}}&	\redbox{\texttt{26}}&    \redbox{\texttt{200}}&		
        \redbox{\texttt{28}}&	\greenbox{\texttt{0}}&	\greybox{\texttt{69}}&	    	
        \greybox{\texttt{11}}&	\redbox{\texttt{26}}&	\redbox{\texttt{200}}&        
        \greenbox{\texttt{4}}&	\greybox{\texttt{18}}&	\greenbox{\texttt{5}}\\
        
S10&    \greybox{\texttt{7}}&	\greenbox{\texttt{0}}&	\redbox{\texttt{705}}&  	    
        \greybox{\texttt{12}}&	\redbox{\texttt{1}}&	    \greybox{\texttt{11}}&		
        \redbox{\texttt{20}}&	\greenbox{\texttt{0}}&	\greybox{\texttt{46}}&	    	
        \greybox{\texttt{12}}&	\redbox{\texttt{1}}&	    \greybox{\texttt{11}}&	    	
        \greenbox{\texttt{5}}&	\greenbox{\texttt{0}}&	\greenbox{\texttt{1}}\\
        
S11&	\greenbox{\texttt{7}}&	\greenbox{\texttt{0}}&	\redbox{\texttt{30}}&		    
        \greenbox{\texttt{7}}&	\redbox{\texttt{2}}&	    \greybox{\texttt{19}}&		
        \redbox{\texttt{15}}&	\greenbox{\texttt{0}}&	\greybox{\texttt{12}}&	    	
        \greenbox{\texttt{7}}&	\redbox{\texttt{2}}&	    \greybox{\texttt{19}}&		    
        \greenbox{\texttt{7}}&	\greenbox{\texttt{0}}&	\greenbox{\texttt{1}}\\
        
S12&	\greenbox{\texttt{7}}&	\greenbox{\texttt{0}}&	\redbox{\texttt{67}}&		    
        \greybox{\texttt{10}}&	\redbox{\texttt{1}}&	    \greybox{\texttt{7}}&		
        \redbox{\texttt{18}}&	\greenbox{\texttt{0}}&	\greybox{\texttt{20}}&	    	
        \greybox{\texttt{10}}&	\redbox{\texttt{1}}&	    \greybox{\texttt{7}}&	    	
        \greenbox{\texttt{7}}&	\greenbox{\texttt{0}}&	\greenbox{\texttt{1}}\\
        
S13&	\greybox{\texttt{10}}&	\greenbox{\texttt{0}}&	\greybox{\texttt{23}}&		    
        \greybox{\texttt{9}}&	\redbox{\texttt{2}}& 	\greybox{\texttt{8}}&		
        \redbox{\texttt{17}}&	\greenbox{\texttt{0}}&	\redbox{\texttt{58}}&	    	
        \greybox{\texttt{9}}&	\redbox{\texttt{2}}&	    \greybox{\texttt{8}}&	    	
        \greenbox{\texttt{5}}&	\greenbox{\texttt{0}}&	\greenbox{\texttt{1}}\\
        
S14&	\greenbox{\texttt{5}}&	\greenbox{\texttt{0}}&	\greybox{\texttt{17}}&		    
        \redbox{\texttt{6}}&	    \redbox{\texttt{1}}&	    \greybox{\texttt{6}}&		
        \redbox{\texttt{6}}&	    \greenbox{\texttt{0}}&	\redbox{\texttt{18}}&	    	
        \redbox{\texttt{6}}&	    \redbox{\texttt{1}}&	    \greybox{\texttt{6}}&	    	
        \redbox{\texttt{6}}&	    \greenbox{\texttt{0}}&	\greenbox{\texttt{1}}\\
        
\rowcolor{LightCyan2}
\textbf{T/A}&    \textbf{\texttt{81}}&   \greenbox{\textbf{\texttt{0}}}&	\redbox{\textbf{\texttt{115.3}}}&	       \textbf{\texttt{134}}&            \redbox{\textbf{\texttt{77}}}&    \textbf{\texttt{46}}&	
        \redbox{\textbf{\texttt{217}}}&	\greenbox{\textbf{\texttt{0}}}&   \textbf{\texttt{47}}&	  	
        \textbf{\texttt{134}}&            \redbox{\textbf{\texttt{77}}}&	\textbf{\texttt{46}}&	        
        \greenbox{\textbf{\texttt{70}}}&	\textbf{\texttt{45}}&	            \greenbox{\textbf{\texttt{1.7}}}\\

\midrule

C1& 	\greenbox{\texttt{8}}&   \greenbox{\texttt{0}}&	\redbox{\texttt{38}}&		    
        \redbox{\texttt{11}}&	\redbox{\texttt{1}}&	    \greybox{\texttt{11}}&		
        \greybox{\texttt{RE}}&	\greybox{\texttt{RE}}&	\greybox{\texttt{RE}}&	    	
        \redbox{\texttt{11}}&	\redbox{\texttt{1}}&	    \greybox{\texttt{11}}&		    
        \greenbox{\texttt{8}}&	\greenbox{\texttt{0}}&	\greenbox{\texttt{1}}\\
        
C2& 	\greenbox{\texttt{8}}&   \greenbox{\texttt{0}}&	\redbox{\texttt{44}}&		    
        \redbox{\texttt{11}}&	\redbox{\texttt{1}}&	    \greybox{\texttt{7}}&		
        \greybox{\texttt{RE}}&	\greybox{\texttt{RE}}&	\greybox{\texttt{RE}}&	     	
        \redbox{\texttt{11}}&	\redbox{\texttt{1}}&	    \greybox{\texttt{7}}&	    	
        \greenbox{\texttt{8}}&	\greenbox{\texttt{0}}&	\greenbox{\texttt{1}}\\
        
C3&	    \redbox{\texttt{30}}&    \greenbox{\texttt{0}}&	\redbox{\texttt{121}}&	    
        \greybox{\texttt{21}}&	\redbox{\texttt{3}}&	    \greybox{\texttt{12}}&		
        \greybox{\texttt{RE}}&	\greybox{\texttt{RE}}&	\greybox{\texttt{RE}}&	    	
        \greybox{\texttt{21}}&	\redbox{\texttt{3}}&	    \greybox{\texttt{12}}&	    	
        \greenbox{\texttt{11}}&	\greenbox{\texttt{0}}&	\greenbox{\texttt{1}}\\
        
C4& 	\greenbox{\texttt{12}}&  \greenbox{\texttt{0}}&	\redbox{\texttt{24}}&		    
        \redbox{\texttt{28}}&	\greenbox{\texttt{0}}&	\greybox{\texttt{3}}&		
        \greybox{\texttt{RE}}&   \greybox{\texttt{RE}}&   \greybox{\texttt{RE}}&         
        \redbox{\texttt{28}}&    \greenbox{\texttt{0}}&	\greybox{\texttt{3}}&	       
        \greybox{\texttt{18}}&   \greenbox{\texttt{0}}&	\greenbox{\texttt{1}}\\
        
C5& 	\greybox{\texttt{16}}&   \greenbox{\texttt{0}}&	\redbox{\texttt{320}}&	    
        \redbox{\texttt{33}}&	\greenbox{\texttt{0}}&	\greybox{\texttt{3}}&		
        \greybox{\texttt{RE}}&   \greybox{\texttt{RE}}&   \greybox{\texttt{RE}}&	    
        \redbox{\texttt{33}}&    \greenbox{\texttt{0}}&	\greybox{\texttt{3}}&		    
        \greenbox{\texttt{10}}&	\greenbox{\texttt{0}}&	\greenbox{\texttt{1}}\\
        
C6& 	\greenbox{\texttt{9}}&   \greenbox{\texttt{0}}&	\redbox{\texttt{311}}&	    
        \redbox{\texttt{24}}&	\greenbox{\texttt{0}}&	\greybox{\texttt{2}}&		
        \greybox{\texttt{RE}}&	\greybox{\texttt{RE}}&	\greybox{\texttt{RE}}&	    	
        \redbox{\texttt{24}}&	\greenbox{\texttt{0}}&	\greybox{\texttt{2}}&	    	
        \greenbox{\texttt{9}}&	\greenbox{\texttt{0}}&	\greenbox{\texttt{1}}\\
        
C7& 	\greenbox{\texttt{9}}&   \greenbox{\texttt{0}}&	\redbox{\texttt{38}}&		    
        \redbox{\texttt{17}}&	\redbox{\texttt{2}}&	    \greybox{\texttt{9}}&		
        \greybox{\texttt{RE}}&	\greybox{\texttt{RE}}&	\greybox{\texttt{RE}}&	    	
        \redbox{\texttt{17}}&	\redbox{\texttt{2}}&	    \greybox{\texttt{9}}&	    	
        \greenbox{\texttt{9}}&	\greenbox{\texttt{0}}&	\greenbox{\texttt{1}}\\
        
C8& 	\greenbox{\texttt{11}}&  \greenbox{\texttt{0}}&	\redbox{\texttt{27}}&		    
        \redbox{\texttt{25}}&	\redbox{\texttt{2}}&	    \greybox{\texttt{11}}&		
        \greybox{\texttt{RE}}&	\greybox{\texttt{RE}}&	\greybox{\texttt{RE}}&	    	
        \redbox{\texttt{25}}&	\redbox{\texttt{2}}&	    \greybox{\texttt{11}}&	    	
        \greenbox{\texttt{11}}&	\greenbox{\texttt{0}}&	\greenbox{\texttt{1}}\\
        
C9&	    \redbox{\texttt{19}}&    \greenbox{\texttt{0}}&	\redbox{\texttt{452}}&	    
        \greybox{\texttt{16}}&	\redbox{\texttt{2}}&	    \greybox{\texttt{17}}&	 	
        \greybox{\texttt{RE}}&	\greybox{\texttt{RE}}&	\greybox{\texttt{RE}}&	    	
        \greybox{\texttt{16}}&	\redbox{\texttt{2}}&	    \greybox{\texttt{17}}&	    	
        \greenbox{\texttt{9}}&	\greenbox{\texttt{0}}&	\greenbox{\texttt{1}}\\
        
C10&	\greybox{\texttt{12}}&   \greenbox{\texttt{0}}&	\redbox{\texttt{142}}&		    
        \redbox{\texttt{13}}&	\greenbox{\texttt{0}}&	\greybox{\texttt{3}}&		
        \greybox{\texttt{RE}}&	\greybox{\texttt{RE}}&	\greybox{\texttt{RE}}&	    	
        \redbox{\texttt{13}}&	\greenbox{\texttt{0}}&	\greybox{\texttt{3}}&	    	
        \greenbox{\texttt{11}}&	\greenbox{\texttt{0}}&	\greenbox{\texttt{1}}\\
        
\rowcolor{LightCyan2}
\textbf{T/A}&    \textbf{\texttt{134}}&\greenbox{\textbf{\texttt{0}}}& \redbox{\textbf{\texttt{151.7}}}&	   
        \redbox{\textbf{\texttt{199}}}&   \redbox{\textbf{\texttt{11}}}&    \textbf{\texttt{7.8}}&	
        \textbf{\texttt{NA}}&             \textbf{\texttt{NA}}&   \textbf{\texttt{NA}}&	    
        \redbox{\textbf{\texttt{199}}}&   \textbf{\texttt{11}}&	\textbf{\texttt{7.8}}&		
        \greenbox{\textbf{\texttt{104}}}&	\greenbox{\textbf{\texttt{0}}}&	\greenbox{\textbf{\texttt{1}}}\\

\midrule

CH1& 	\greenbox{\texttt{22}}&	        \greenbox{\texttt{0}}&	\redbox{\texttt{333}}&		    
        \redbox{\texttt{41}}&	        \greybox{\texttt{48}}&	\greybox{\texttt{62}}&		
        \greybox{\texttt{RE}}&	\greybox{\texttt{RE}}&	\greybox{\texttt{RE}}&	     	
        \redbox{\texttt{41}}&	        \greybox{\texttt{48}}&	\greybox{\texttt{62}}&	    	
        \greenbox{\texttt{22}}&	        \greenbox{\texttt{0}}&	\greenbox{\texttt{3}}\\
        
CH2&	\greenbox{\texttt{10}}&	\greenbox{\texttt{0}}&	\redbox{\texttt{196}}&	        
        \redbox{\texttt{20}}&	\redbox{\texttt{32}}&	\greybox{\texttt{96}}&		
        \greybox{\texttt{RE}}&	\greybox{\texttt{RE}}&	\greybox{\texttt{RE}}&	    	
        \redbox{\texttt{20}}&	\redbox{\texttt{32}}&	\greybox{\texttt{96}}&	    	
        \greenbox{\texttt{10}}&	\greenbox{\texttt{0}}&	\greenbox{\texttt{5}}\\
        
CH3& 	\greybox{\texttt{18}}&	\greenbox{\texttt{0}}&	\greybox{\texttt{544}}&		    
        \redbox{\texttt{37}}&    \redbox{\texttt{37}}&	\redbox{\texttt{604}}&		
        \greybox{\texttt{RE}}&   \greybox{\texttt{RE}}&   \greybox{\texttt{RE}}&         
        \redbox{\texttt{37}}&    \redbox{\texttt{37}}&	\redbox{\texttt{604}}&	    
        \greenbox{\texttt{13}}&  \greenbox{\texttt{0}}&	\greenbox{\texttt{4}}\\
        
CH4& 	\greybox{\texttt{RE}}&	\greybox{\texttt{RE}}&	\greybox{\texttt{RE}}&	        
        \redbox{\texttt{18}}&    \redbox{\texttt{28}}&	\redbox{\texttt{25}}&		
        \greybox{\texttt{RE}}&   \greybox{\texttt{RE}}&   \greybox{\texttt{RE}}&	    
        \redbox{\texttt{18}}&    \redbox{\texttt{28}}&	\redbox{\texttt{25}}&		    
        \greenbox{\texttt{12}}&	\greenbox{\texttt{0}}&	\greenbox{\texttt{3}}\\
        
CH5& 	\greenbox{\texttt{11}}&	\greenbox{\texttt{0}}&	\redbox{\texttt{522}}&	 
        \redbox{\texttt{29}}&	\redbox{\texttt{41}}&  \greenbox{\texttt{48}}&
        \greybox{\texttt{RE}}&	\greybox{\texttt{RE}}&	\greybox{\texttt{RE}}&	    	
        \redbox{\texttt{29}}&	\redbox{\texttt{41}}&  \greenbox{\texttt{48}}&	        	    	
        \greybox{\texttt{RE}}&	\greybox{\texttt{RE}}&	\greybox{\texttt{RE}}\\
        
CH6& 	\greenbox{\texttt{15}}&	\greenbox{\texttt{0}}&	\redbox{\texttt{311}}&		    
        \redbox{\texttt{34}}&	\redbox{\texttt{54}}&	\greenbox{\texttt{42}}&  
        \greybox{\texttt{RE}}&	\greybox{\texttt{RE}}&	\greybox{\texttt{RE}}&	    	
       \greybox{\texttt{RE}}&	\greybox{\texttt{RE}}&	\greybox{\texttt{RE}}&		    	
        \greybox{\texttt{RE}}&	\greybox{\texttt{RE}}&	\greybox{\texttt{RE}}\\
        
CH7& 	\greenbox{\texttt{26}}&	\greenbox{\texttt{0}}&	\redbox{\texttt{337}}&		    
        \redbox{\texttt{47}}&	\redbox{\texttt{65}}&  \greybox{\texttt{57}}&	    	
        \greybox{\texttt{RE}}&	\greybox{\texttt{RE}}&	\greybox{\texttt{RE}}&	    	
        \redbox{\texttt{47}}&	\redbox{\texttt{65}}&  \greybox{\texttt{57}}&		
        \greenbox{\texttt{26}}&	\greenbox{\texttt{0}}&	\greenbox{\texttt{7}}\\
        
CH8&	\greenbox{\texttt{35}}&	\greenbox{\texttt{0}}&	\redbox{\texttt{126}}&	        
        \redbox{\texttt{36}}&	\redbox{\texttt{77}}&  \greybox{\texttt{66}}&	 	
        \greybox{\texttt{RE}}&	\greybox{\texttt{RE}}&	\greybox{\texttt{RE}}&	    	
        \redbox{\texttt{36}}&	\redbox{\texttt{77}}&  \greybox{\texttt{66}}&	    	
        \greenbox{\texttt{35}}&	\greenbox{\texttt{0}}&	\greenbox{\texttt{6}}\\

\rowcolor{LightCyan2}
\textbf{T/A}&    \textbf{\texttt{137}}& \greenbox{\textbf{\texttt{0}}}& \redbox{\textbf{\texttt{338.4}}}&
        \redbox{\textbf{\texttt{262}}}&  \redbox{\textbf{\texttt{382}}}& \textbf{\texttt{125}}&	    
        \textbf{\texttt{NA}}&             \textbf{\texttt{NA}}&             \textbf{\texttt{NA}}&	    
        \textbf{\texttt{228}}&          \textbf{\texttt{328}}&  \textbf{\texttt{136}}&		
        \greenbox{\textbf{\texttt{98}}}&\greenbox{\textbf{\texttt{0}}}& \greenbox{\textbf{\texttt{4.6}}}\\

\bottomrule
\end{tabular}
\label{tab:metric2}
\end{table*}

 \subsection{Indexing and Source Selection Metrics}
 A smaller-sized index is essential for fast index lookup during source selection, but it can lack important information. In contrast, large index sizes provide slow index lookup and are hard to manage, but may lead to better cardinality estimations. To this end, it is important to compare the size of the indexes generated by the selected federation engines. Table \ref{tab:indx} shows a comparison of the index/data summaries' construction time and the index size\footnote{The index size is given by size of summaries used for cardinality estimation (in MBs).} of the selected state-of-the-art cost-based SPARQL federation approaches. SemaGrow, SPLENDID and LHD rely on VOID statistics with a size of 1 MB for the complete LargeRDFBench datasets of size 34.3 GB.  CostFed's index  size  is 10.5 MB while Odyssey's is 5.2 GB. The much bigger index size used by Odyssey might makes this approach less appropriate to be used for Big RDF datasets such as WikiData, Linked Geo Data etc. CostFed’s index construction time is around 1 hr and 6 mins for the complete LargeRDFBench datasets. SPLENDID, SemaGrow and LHD took 1 hr and 50 mins to generate the index. The Index construction time for Odyssey was 86 hrs and 30 mins, which makes it difficult to use for big datasets or datasets with frequent updates. 
 
 According to \cite{LargeRDFBench2018}, the efficiency of source selection can be measured in terms of: (1) total number of triple pattern-wise sources selected (\#T), (2) the number of SPARQL ASK requests sent to the endpoints (\#A) during source selection, and (3) the source selection time. Table \ref{tab:metric2} shows a comparison of the source selection algorithms of the select triple stores across these metrics. As discussed previously, the smaller \#T leads to better query plan generation \cite{LargeRDFBench2018}. The smaller \#A leads to smaller source selection time, which in turn leads to smaller 
 query execution time. 
 In this regard, CostFed ranked first (83 green boxes, i.e., the best results among the selected engines), followed by Odyssey with 56 green boxes, LHD with 15 green boxes, SPLENDID with 10 green boxes, and then SemaGrow with 9 green boxes. 
 
 The approaches that perform a join-aware and hybrid (SPARQL + index) source selection lead to smaller \#T \cite{LargeRDFBench2018}. Both Odyssey and Costfed perform join-aware source selection and hence lead to smaller \#T than other selected approaches. The highest number of SPARQL ASK requests is sent by index-free federation engines, followed by hybrid (SPARQL + index), which in turn is followed by index-only federation engines \cite{LargeRDFBench2018}. This is because for index-free federation engines, such as FedX, the complete source selection is based on SPARQL ASK queries. The Hybrid engines such as CostFed, SPLENDID, SemaGrow and Odyssey make use of both index and SPARQL ASK queries to perform source selection, thus some of the SPARQL ASK requests are skipped due to the information used in the index. The Index-only engines, such as LHD, only make use of the index to perform the complete source selection. Thus, these engines do not consume a single SPARQL ASK query during source selection. The source selection time for such engines is much smaller due to only index-lookup without sending outside requests to endpoints. 
 However, they have more \#T than hybrid (SPARQL + index) source selection approaches. 

\subsection{Query Execution Time}
\begin{figure*}[!htb]
\centering
\begin{subfigure}{0.98\textwidth}
\begin{adjustwidth}{-1cm}{}
  \centering
  \includegraphics[width=\textwidth]{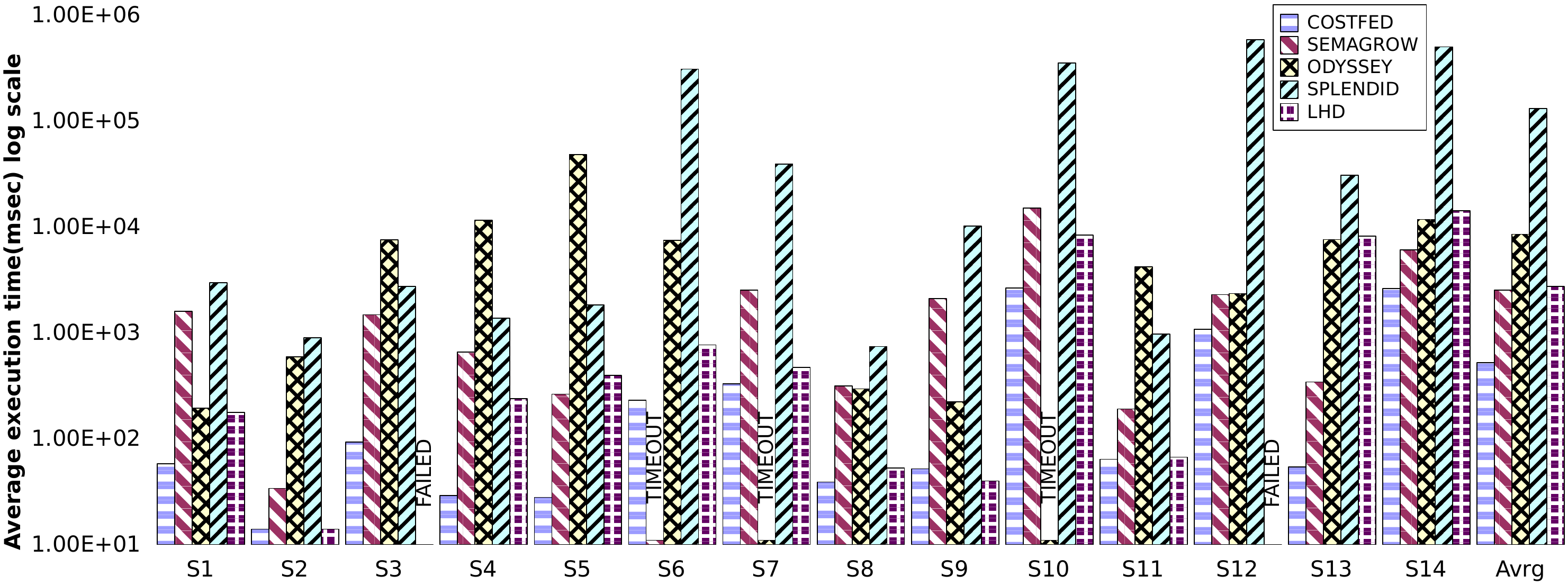}
  \caption{Average execution time of simple (S) queries (FedBench)}
  \label{Figure:SQueriesRuntime}
   \end{adjustwidth}

\end{subfigure}
\begin{subfigure}{0.98\textwidth}
\begin{adjustwidth}{-1cm}{}
  \centering
  \includegraphics[width=\textwidth]{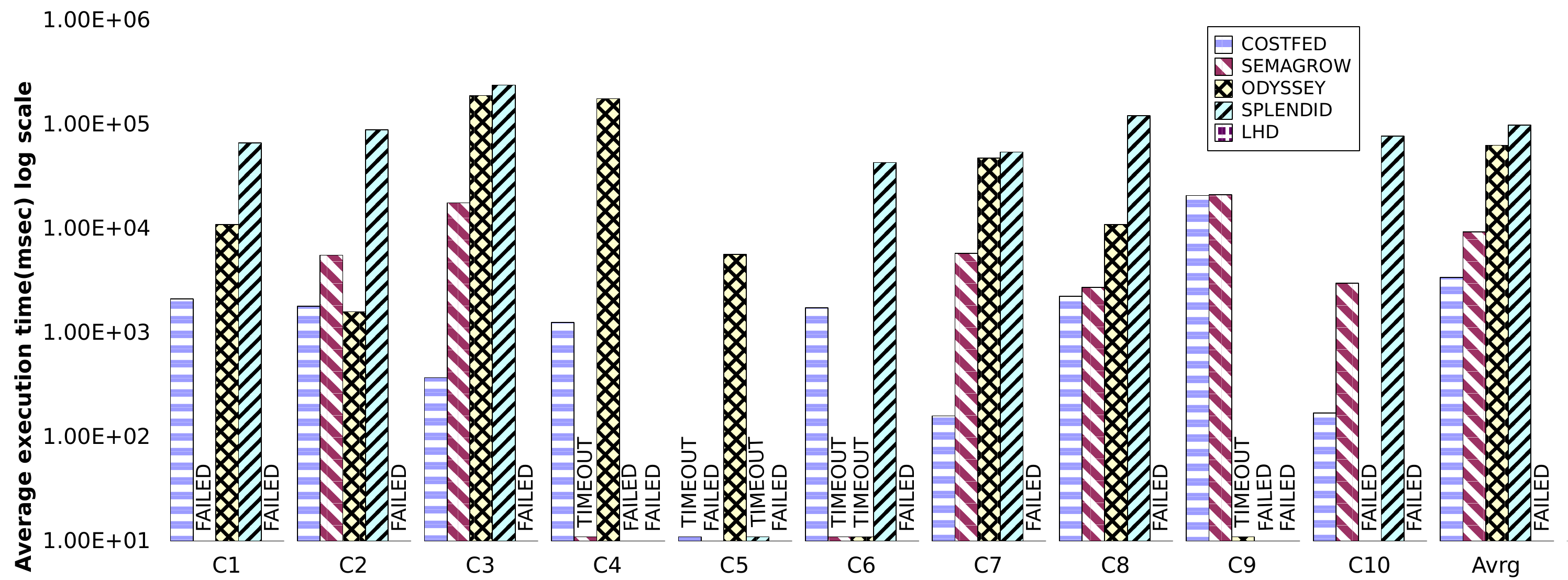}
  \caption{Average execution time of complex (C) queries (LargeRDFBench)}
  \label{Figure:CQueriesRuntime}
   \end{adjustwidth}

\end{subfigure}
\begin{subfigure}{0.98\textwidth}
\begin{adjustwidth}{-1cm}{}
  \centering
  \includegraphics[width=\textwidth]{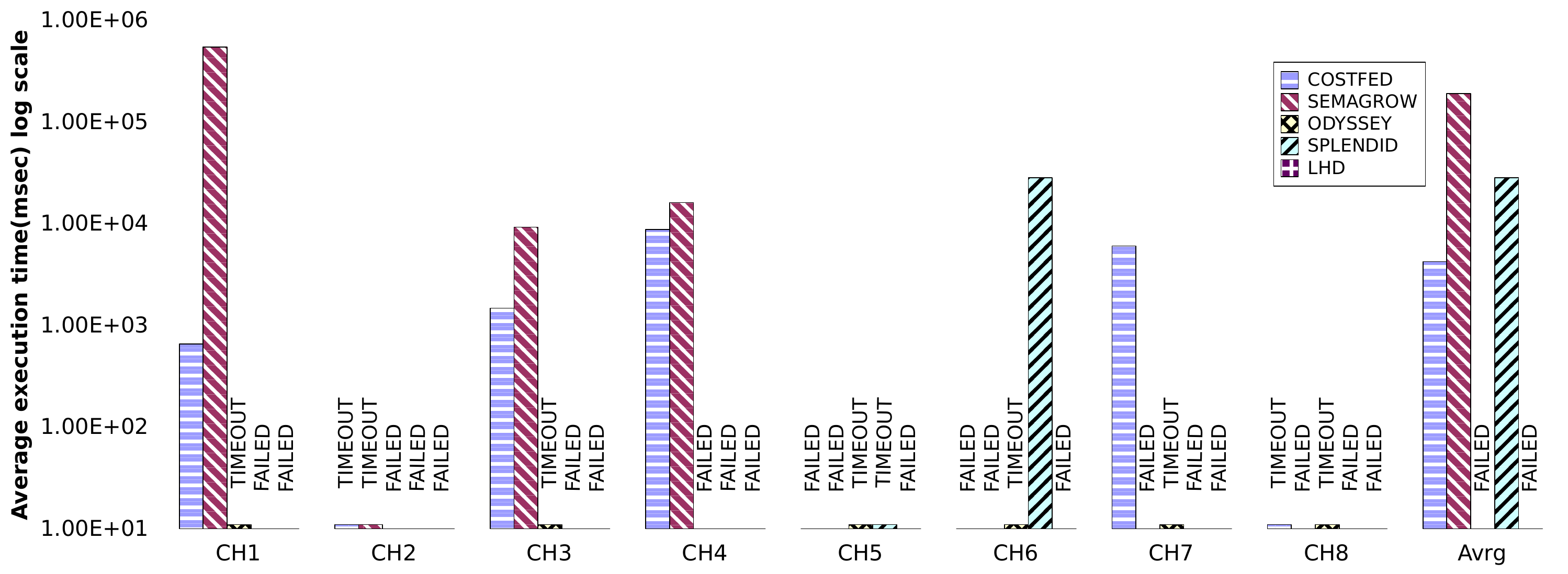}
  \caption{Average execution time of complex and high data structures (ch) queries(LargeRDFBench)}
  \label{Figure:chQueriesRuntime}
   \end{adjustwidth}

\end{subfigure}
\caption{Average execution time of LargeRDFBench and FedBench Queries. }
\label{fig:b1-b2_srcruntime}
\end{figure*}

Finally, we present the query runtime results of the selected federation engines across the different queries categories of LargeRDFBench. 
Figure \ref{fig:b1-b2_srcruntime} gives an overview of our results. In our runtime evaluation on simple queries (S1-S14) (see Figure \ref{Figure:SQueriesRuntime}), CostFed has the shortest runtimes, followed by SemaGrow, LHD, Odyssey, and SPLENDID. CostFed's runtimes are shorter than SemaGrow's on 13/13 comparable queries (excluding queries with timeout and runtime error) (average runtime = 0.5 sec for CostFed vs. 2.5 sec for SemaGrow). SemaGrow outperforms LHD on 4/11 comparable queries with an average runtime of 2.5 sec for SemaGrow vs. 2.7 sec for LHD. LHD's runtimes are shorter than Odyssey's on 8/10 comparable queries with an average runtime of 8.5 sec for Odyssey. Finally, Odyssey is clearly faster than SPLENDID on 8/12 comparable queries with an average runtime of 131 sec for SPLENDID. 

Our runtime evaluation on the complex queries (C1-C10) (see Figure \ref{Figure:CQueriesRuntime}) leads to a different ranking: CostFed produces the shortest runtimes followed by SemaGrow, Odyssey, and SPLENDID. CostFed outperforms SemaGrow in 6/6 comparable queries (excluding queries with timeout and runtime error) with an average runtime of 3 sec for CostFed vs. 9 sec for SemaGrow. SemaGrow's runtimes are shorter than Odyssey's in 3/4 comparable queries with an average runtime of 63 sec for Odyssey. Odyssey is better than SPLENDID in 5/5 comparable queries, where SPLENDID's average runtime is 98 sec.  

The runtime evaluation on the complex and high sources queries (CH1-C8) given in Figure \ref{Figure:chQueriesRuntime} establishes CostFed as the best query federation engine, followed by SPLENDID and then SemaGrow. CostFed's runtimes are smaller than SemaGrow in 3/3 comparable queries (excluding queries with timeout and runtime error), with an average runtime of 4 sec for CostFed vs. 191 sec for SemaGrow. SPLENDID has no comparable queries with CostFed and SemaGrow. LHD and Odyssey both fail to produce results when faced with complex queries.

\section{Conclusion}
\label{sec:conclusion}
In this paper, we presented an extensive evaluation of existing cost-based federated query engines. We used existing metrics from relational 
database research and proposed new metrics to measure the quality of  cardinality estimators of selected engines. To the best of our knowledge, this work is  the first evaluation of cost-based SPARQL federation engines focused on the quality of the cardinality estimations. 

\begin{itemize}
    \item The proposed similarity-based errors have a more positive correlation with runtimes, i.e., the smaller the error values, the better the query runtimes. Thus, this metric helps developers to design a more efficient query execution planner for federation engines. Our proposed approach produces more significant results compared to q-error. However, there is still room for further improvement. 
    \item The higher coefficients (R values) with similarity errors (as opposed to q-error), suggest that the proposed similarity errors are a better predictor for runtime than the q-error. 
    \item The smaller p-values of the similarity errors, as compared to q-error, further confirm that similarity errors are more likely to be a better predictor for runtime than the q-error.
    \item Errors in the cardinality estimation of triple patterns have a higher correlation to runtimes than the error in the cardinality estimation of joins. 
    Thus, cost-based federation engines must pay particular attention to attaining accurate cardinality estimations of triple patterns. 
    \item The number of transferred tuples have a direct co-relation with query runtime, i.e., the smaller the number of transferred tuples, the smaller the query runtimes. 
    \item The smaller number of triple pattern-wise sources selected is key to generate maximum only possible query plans (\texttt{only-plan}). 
    \item On average, the CostFed engine produces the fewest estimation errors and has the shortest execution time for the majority of LargeRDFBench queries. 
    \item The weak to moderate correlation of the cardinality errors with query execution time suggests that the query runtime is a complex measure affected by multi-dimensional performance metrics and SPARQL query features. The proposed similarity error metric is related to the query planning component of the federation engines and is useful for evaluating the quality of the query plans generated by these engines.
    \item The proposed cardinality estimating metrics are generic and can be applied to non-federated cardinality-based query processing engines as well.

\end{itemize}
The impact of our proposed work is to provide new measures for the development of better cost-based federated SPARQL query engines. Furthermore, our proposed metrics will help in determining the quality of the generated query plans, such as indicating whether or not the join orders are correct. This kind of information is not revealed from the query runtime because the overall query runtime is affected by all metrics given in Table \ref{tab:metric-stats}.  
As future work, we want to compare heuristic-based (index-free) federated SPARQL query processing engines with cost-based federated engines. We want to investigate how much an index is assisting a cost-based federated SPARQL engine to generate optimized query execution plans.   
\section*{Acknowledgments}
The work has been supported by the EU H2020 Marie Skłodowska-Curie project KnowGraphs (no. \ 860801), BMVI-funded project LIMBO (Grant no.\ 19F2029I), BMVI-funded project OPAL (no.\ 19F2028A), and BMBF-funded EuroStars project SOLIDE (no. 13N14456). This work has also been supported by the National Research Foundation of Korea (NRF) (grant funded by the Korea government (MSIT) (no. NRF-2018R1A2A2A05023669)). 

\bibliographystyle{plain}
\bibliography{cost_based}
\end{document}